\documentclass[prc,showpacs,floatfix,aps,10pt]{revtex4-2}
\usepackage{graphicx}% Include figure files
\usepackage{bm}% bold math
\begin{document}

%%%%%%%%%%%%%%%%%%
%%% Title Page %%%
%%%%%%%%%%%%%%%%%%
\title{
Low-energy ${}^{12}$C continuum states in three-$\alpha$ model
}
\author{Souichi Ishikawa}  \email[E-mail:]{ishikawa@hosei.ac.jp}
\affiliation{Science Research Center, Hosei University, 
2-17-1 Fujimi, Chiyoda, Tokyo 102-8160, Japan} 

\date{\today}

\begin{abstract}
The electric multipole strength distributions for transitions from the ${}^{12}\mathrm{C}(0_1^+)$ ground state to $3\alpha$ ($0^+$, $1^-$, $2^+$, and $3^-$) continuum states are studied in terms of $3\alpha$ model. 
Several sets of the $3\alpha$ Hamiltonian are introduced phenomenologically with conventional $\alpha$-$\alpha$ interaction potentials and  $3\alpha$ potentials with different range parameters. 
The transition strength distributions are obtained from wave functions of $3\alpha$ bound- and continuum states calculated by the Faddeev three-body formalism. 
From these strength distributions,  the resonance parameters (energy,  $3\alpha$-decay width, and transition strength) of $3\alpha$ resonance states are extracted. 
Dependencies of these observables on interaction potential models are studied to examine the introduced interaction models. 
Using the transition strength distributions, excitation-energy spectra in the ${}^{12}\mathrm{C}(\alpha,\alpha^\prime)3\alpha$ reaction are calculated to compare with experimental data.
\end{abstract}

\maketitle

%%%%%%%%%%%%%%
%%% Sec. 1 %%%
%%%%%%%%%%%%%%
\section{Introduction}
%%%%%%%%%%%%%%%%

A detailed knowledge of low-energy  excited states in the ${}^{12}$C nucleus plays an important role in studies of various nuclear models and also of astrophysical issues, such as the triple-$\alpha$ process, in which three $\alpha$-particles  in continuum states are fused into a ${}^{12}$C nucleus in stars \cite{Fr14,To17,Fr18}. 
In spite of such importance, there still remain some uncertainties in knowledge of continuum excited states of ${}^{12}$C:  
excitation energies, angular momenta, decay widths,  etc.
In Table \ref{tab:12C-levels}, a summary of energy levels of ${}^{12}$C is shown in comparing the latest three compilation articles \cite{Aj85,Aj90,Ke17}.
Here, the excitation energy $E_x$ is measured from the ground state energy of ${}^{12}$C and the $3\alpha$ energy in the center of mass (c.m.) frame $E$ from the $3\alpha$ breakup threshold.

At energies above the $3\alpha$ breakup threshold ($E>0$), besides the well known resonance states, namely, $0_2^{+}$ state ($E=0.38$ MeV, the Hoyle state), $3_1^-$ state ($E=2.37$ MeV), and the $1_1^-$  state ($E=3.57$ MeV), there remain some unestablished states. 
Actually, the state at $E=3.0$ MeV with a large decay width has been listed in the compilations with temporarily assigned as $0^+$ state for a long time, but  has not been confirmed. 
In the latest compilation \cite{Ke17} after some experimental works  \cite{Fr09,It11,Zi11,Fr12,Zi13}, it is noted that a $2^{+}$ state at $E=2.56$ MeV and a $0^{+}$ state at $E=2.66$ MeV can be possible components for the $E=3.0$ MeV state. 
Also in Ref. \cite{Ke17}, it was noted  that the $0^{+}$ state at $E=2.66$ MeV might be a doublet of $E=1.77$ MeV and $E=3.29$ MeV states according to an experimental analysis of the ${}^{12}\mathrm{C}(\alpha,\alpha^\prime)3\alpha$ reaction \cite{It11}.
This doublet structure in the low-energy $0^+$ state was confirmed in  recent analysis of the inclusive excitation-energy spectra in  ${}^{12}\mathrm{C}(\alpha,\alpha^\prime){}^{12}\mathrm{C}$ and ${}^{14}\mathrm{C}(p,t){}^{12}\mathrm{C}$ reactions in terms of a multi-level, multi-channel R-matrix theory \cite{Li22a,Li22b}.

Theoretically, the existence of $0^{+}$ state in singlet or in doublet as well as $2^{+}$ state around $E=3$ MeV has been reported in semi-microscopic $3\alpha$ cluster model calculations  \cite{Ku05,Ku07,Oh13},  microscopic $3\alpha$ cluster model calculations  \cite{Zh16,Fu15,Fu16}, and microscopic  nuclear calculations \cite{Ka07,Ch07,Ep12}.
These calculations are based on bound state approximations, in which continuous energy eigenvalues for $E>0$  are discretized.  

On the other hand, the present author has performed rather simple $3\alpha$ model calculations without applying any bound state approximation \cite{Is13,Is14a,Is14b,Is16}.
There, $\alpha$-particles are considered as $0^+$ structureless particles, and  some phenomenological sets of $2\alpha$- and $3\alpha$ interaction potentials are introduced to construct $3\alpha$ Hamiltonians, whose eigenstates consist of two bound states, ${}^{12}\mathrm{C}(0_1^+)$ and ${}^{12}\mathrm{C}(2_1^+)$, and $3\alpha$  continuum states with $E>0$. 
In Refs. \cite{Is13,Is14a,Is14b,Is16}, strength distributions for transitions from the $3\alpha$ bound states to $3\alpha$-$0^+$ and -$2^+$ continuum states were calculated, which reveal two-peak structure in $0^+$ state and a peak in $2^+$ state around $E=3$ MeV. 
In general, a peak appeared in the strength distribution as a function of $E$ may correspond to a resonance state of the system, and the peaks appeared in the $3\alpha$ model calculations are recognized as the resonance states mentioned above. 

The interaction potentials among $\alpha$-particles used in the calculations include some phenomenological parameters representing their ranges and strengths, which can not be determined {\it a priori}.
Although these parameters are determined to reproduce some limited numbers of established $2\alpha$ and $3\alpha$ experimental data,  uncertainties can not be avoided.   
Actually, the heights of the peaks and thus the line shapes of the strength distributions strongly depend on the choice of potential parameters

In the present paper, I will study the interaction model dependence of the electric multipole (E$\lambda$) transition strength distributions up to $\lambda=3$ by calculating $0^+$-, $1^-$-, $2^+$-, and $3^-$ $3\alpha$ continuum states adapting the $3\alpha$ model. 
These transition strength distributions are important inputs to calculate some reactions whose final states include $3\alpha$ breakup states. 
As a typical example of such reactions, inelastic scattering of $\alpha$-particle from ${}^{12}$C,  ${}^{12}\mathrm{C}(\alpha,\alpha^\prime)3\alpha$ reaction, will be calculated by adapting a formalism of plane-wave impulse approximation (PWIA) as the reaction mechanism, and experimental data of the excitation-energy spectra in Ref. \cite{It11} will be analyzed.

In Sec. \ref{sec:formalism}, the formalism to calculate the transition strength distributions with $3\alpha$ continuum states  will be shortly summarized. 
Sec. \ref{sec:strength_distributions}, after introducing interaction potential models, I will show results of the strength distributions for $\lambda=0$ to 3 multipoles. 
The calculated excitation-energy spectra in ${}^{12}{\rm C}(\alpha,\alpha^\prime)3\alpha$ reaction comparing with data of Ref. \cite{It11} will be presented in Sec. \ref{sec:energy_spectrum}.
Summary will be given in Sec. \ref{sec:summary}.

%%%%%%%%%%%%%%%%%%%%%%%%%%%%%%%%%%%%%%%
%%% Table I: Summary of 12C energy level
%%%%%%%%%%%%%%%%%%%%%%%%%%%%%%%%%%%%%%%
\begin{table}[t]
\caption{
Summary of energy levels of  ${}^{12}$C nucleus in the compilations \cite{Aj85,Aj90,Ke17}. 
Only  $0^+$, $1^-$, $2^+$, and $3^-$ with isospin 0 states are shown.
Excitation energy $E_x$, angular momentum $J$, parity $\pi$, energy measured from the $3\alpha$ breakup threshold $E$, and decay width $\Gamma$, which are used in this paper, are shown. 
Numbers, some of which are rounded off adequately, are taken from the latest compilation.
Numbers enclosed in parentheses indicate to include uncertainty. 
\label{tab:12C-levels} 
}
\begin{ruledtabular}
\begin{tabular}{cccc}
$E_x$   &   $J^\pi$ & $E$  &  $\Gamma$  \\   
 MeV $\pm$ keV  &  &   MeV  &   keV \\ 
\hline
0    &   $0^{+}$   & -7.2746   &  \\  
4.43982 $\pm$ 0.21 &  $2^{+}$  & -2.8348   &  \\  
7.65407 $\pm$ 0.19    &   $0^{+}$   &  0.3795    & 9.3 $\pm$ 0.9 eV \\ 
9.641 $\pm$ 5    &   $3^{-}$   & 2.367   & 46 $\pm$ 3 \footnote{34 $\pm$ 5 keV in Refs. \cite{Aj85,Aj90}.}\\ 
9.870 $\pm$ 60 \footnote{In Ref. \cite{Ke17}, but not in Refs. \cite{Aj85,Aj90}.}    &   $2^{+}$   & 2.595   & 850 $\pm$ 85 \\  
(9.930  $\pm$ 30)  \footnote{In Ref. \cite{Ke17}, but not in Refs. \cite{Aj85,Aj90}.}  &   $0^{+}$   & 2.655   &  2710 $\pm$ 80\\  
10.3  $\pm$ 300  \footnote{In Refs. \cite{Ke17,Aj85,Aj90}, but with some uncertainties.}   &   $(0^{+})$   & 3.0   & 3000 $\pm$ 700 \\  
10.847  $\pm$ 4  &   $1^{-}$   & 3.572   & 273 $\pm$ 5  \\  
11.16 $\pm$ 500  \footnote{In Refs. \cite{Aj85,Aj90}, but not in Ref. \cite{Ke17}.} & $(2^{+})$ &3.89 & 430 $\pm$ 80\\
(15.44 $\pm$ 40)  &   $(2^{+})$   & 8.17   & 1770$\pm$200 \\ 
(18.6 $\pm$ 100) &  $(3^{-})$\footnote{Isospin is not specified.}  & 11.3 & 300 \\
\end{tabular}
\end{ruledtabular}
\end{table}
%%%%%%%%%%%%%%%%%%%%%%%%%%%%%%%%%%%%%%%%%%%%%%%%%%%%%%%%%%%%

%%%%%%%%%%%%%%%%
\section{Formalism}
\label{sec:formalism}
%%%%%%%%%%%%%%%%%%

%%%%%%%%%%%%%%%%
\subsection{$3\alpha$ system}
%%%%%%%%%%%%%%%

Here, I introduce a Hamiltonian of three $\alpha$-particles in the c.m. frame,
\begin{equation}
\hat{H} _{3\alpha} = \hat{H}_0 + \hat{V},
\end{equation} 
where $\hat{H}_0$ is the kinetic energy operator of the $3\alpha$ system, and $\hat{V}$ is an interaction potential, which consists of two-$\alpha$ potentials ($2\alpha$Ps) and three-$\alpha$ potentials ($3\alpha$Ps).
Explicit forms of these potentials will be presented in Sec. \ref{sec:strength_distributions}.

Since there is no $2\alpha$ bound state, two-body incoming scattering experiment is not available for the $3\alpha$ system.
Thus, for studies of  $3\alpha$ continuum states, a transition from a $3\alpha$ bound state $\vert  \Psi_b \rangle$  with energy $E_b(<0)$: 
\begin{equation}
\hat{H}_{3\alpha} \vert  \Psi_b \rangle = E_b \vert  \Psi_b \rangle,
\label{eq:3a_bound_state}
\end{equation} 
to $3\alpha$ continuum states with energy $E(>0)$ will be considered.
The final $3\alpha$ states are specified by two asymptotic momenta (wave numbers): the relative momentum between two $\alpha$-particles, $\bm{q}$, and the momentum of the third $\alpha$-particle with respect to the c.m. of the $\alpha$-particle-pair, $\bm{p}$: 
\begin{equation}
\bm{q} = \frac{1}{2} \left( \bm{k}_1 - \bm{k}_2 \right),
\quad\quad
\bm{p} = \frac{2}{3}\bm{k}_3   - \frac{1}{3}\left(\bm{k}_1 + \bm{k}_2\right) =  \bm{k}_3,
\label{eq:qp-k_i}
\end{equation}
where $\bm{k}_i$ is the momentum of the $i$-th $\alpha$-particle in the $3\alpha$ c.m. frame. 
These variables are not independent but satisfy 
\begin{equation}
E  = \frac{\hbar^2}{m_\alpha} q^2  + \frac{3 \hbar^2}{4 m_\alpha}  p^2,
\label{eq:EqEp}
\end{equation}
where  $m_\alpha$ is the mass of the $\alpha$-particle.

The strength distribution function for the transition process induced by an operator $\hat{O}$ is defined by 
\begin{equation}
S(E)  = 
 \int  d\bm{q}^\prime d\bm{p}^\prime 
\left\vert \langle \Psi^{(-)}_{\bm{q}^\prime, \bm{p}^\prime}(E^\prime) \vert \hat{O} \vert  \Psi_b  \rangle  \right\vert^2
 \delta\left(E- E^\prime \right).
\label{eq:S_strength}
\end{equation}
Here, $\vert \Psi^{(\pm)}_{\bm{q},\bm{p}}(E)\rangle$ is a $3\alpha$ scattering (continuum) state with the initial state specified by $\{ \bm{q},\bm{p} \}$, 
\begin{equation}
\hat{H}_{3\alpha} \vert   \Psi^{(\pm)}_{\bm{q},\bm{p}} (E)\rangle
=E \vert  \Psi^{(\pm)}_{\bm{q},\bm{p}} (E)\rangle, 
\end{equation}
where the superscript $(+)$ [($-$)] expresses the boundary condition with an outgoing  (incoming) $3\alpha$ asymptotic wave.   

Here, I introduce a wave function $ \Xi (\bm{x} , \bm{y})$ describing the transition process as in Ref. \cite{Is94},  
\begin{equation}
\Xi (\bm{x} , \bm{y}) = \langle \bm{x} , \bm{y}   \vert \frac{1}{E+ i \epsilon - \hat{H}_{3\alpha}} \hat{O} \vert \Psi_b \rangle, 
\label{eq:Psi_def}
\end{equation} 
where Jacobi coordinates $\{\bm{x}, \bm{y}\}$  are defined by  
\begin{equation}
\bm{x} =  \bm{r}_1 - \bm{r}_2, 
\quad\quad
\bm{y}  =  \bm{r}_3    - \frac12 \left(\bm{r}_1 + \bm{r}_2 \right),
\label{eq:Jacobi}
\end{equation}
with $\bm{r}_i$ being the coordinate vector of the $i$-th $\alpha$-particle in the $3\alpha$ c.m. frame.
Note that the coordinates,  $\bm{x}$  and $\bm{y}$, are conjugate to the momenta,  $\bm{q}$ and $\bm{p}$, respectively.

When Eq. (\ref{eq:Psi_def}) is solved, the amplitude $\langle \Psi^{(-)}_{\bm{q}, \bm{p}}(E) \vert \hat{O} \vert  \Psi_b  \rangle$ in Eq. (\ref{eq:S_strength}) 
is extracted from the asymptotic form of $\Xi (\bm{x} , \bm{y})$ evaluated by the saddle-point approximation \cite{Sa77}, 
\begin{equation}
\Xi(\bm{x},\bm{y}) 
\displaystyle{\mathop{\to}_{\begin{array}{c} x \to \infty \\ y/x~{\text{fixed}}\end{array}} }
e^{\frac{\pi}{4} i} \sqrt{\frac{\pi}{2}}   \frac{m_\alpha}{\hbar^2}  \left( \frac{4K_0}{3} \right)^{3/2} 
\frac{e^{i \left( K_0 + {\cal O}(R^{-1}) \right)R }}{R^{5/2}} 
\langle \Psi^{(-)}_{q\hat{\bm{x}}, p\hat{\bm{y}}}(E) \vert \hat{O} \vert  \Psi_b  \rangle, 
\label{eq:psi_asym}
\end{equation}
where long-range terms due to the Coulomb interaction are expressed just by ${\cal O}(R^{-1})$  for simplicity. 
Here, the hyper radius $R$ and a momentum $K_0$ are defined by  
\begin{equation}
R = \sqrt{x^2 + \frac43 y^2},
\end{equation}
\begin{equation}
K_0 = \sqrt{ \frac{m_\alpha}{\hbar^2} E},
\end{equation} 
and  $q$ and $p$ satisfy Eq. (\ref{eq:EqEp}) and $\frac{p}{q} = \frac{4y}{3x}$. 

Numerical solutions of Eq. (\ref{eq:Psi_def}) as well as those of bound state problem, Eq. (\ref{eq:3a_bound_state}), are obtained by the method based on the Faddeev three-body theory \cite{Fa61}, in which the equations are expressed as coupled integral equations in the coordinate space, and are solved by an iterative method.
It is remarked that the Faddeev formalism is not directly adaptable to the $3\alpha$ systems due to the presence of the long range Coulomb forces.
A brief description of how  Coulomb effects are treated is given  Appendix \ref{sec:Coul_mod_Faddeev}.
In the present calculations, wave functions for the $3\alpha$ bound (continuum) states are calculated up to 100 fm (1000 fm) for both of  the $x$ and $y$ variables.
Thus matrix elements of the operators between bound and continuum states are calculated up to 100 fm.
Formal and technical details are essentially same as those used for the three-nucleon bound state \cite{Sa86}, and the proton-deuteron scattering \cite{Is03,Is09}, and details of the application of the method to the $3\alpha$ systems are found in Ref. \cite{Is13}.

%%%%%%%%%%%%%
\subsection{Electric multipole strength distribution}
\label{subsection:MSD}
%%%%%%%%%%%%%%%%%%

The electric transition strength is defined in terms of  an electric operator for the ${}^{12}$C nucleus in nucleonic base, 
\begin{equation}
\hat{O}_{\mathrm{E}}^{[12]} = \sum_{k=1,12} \frac{e}{2} \left( 1 + \tau_z(k) \right) F(\bm{s}_k),
\end{equation}
where $\bm{s}_k$  is the coordinate of $k$-th nucleon in the  ${}^{12}$C-c.m. system, $\tau_z(k)$ is the third component  of isospin operator for  $k$-th nucleon, and $F(\bm{s})$ is a function describing the multipole component. 
In calculating a matrix element of this operator in the $3\alpha$ model, one may consider the four nucleons,  $4(i-1)+1,.., 4(i-1)+4, (i=1,2,3)$, are in the same position $\bm{r}_i$ of  $i$-th $\alpha$-cluster consisting of these nucleons. 
Then, using the fact that the total isospin of  the $\alpha$-particle is 0, 
\begin{equation}
\hat{O}_{\mathrm{E}} = 2e \sum_{i=1,3}  F(\bm{r}_i)
\end{equation} 
is obtained for the effective operator of $\hat{O}_{\mathrm{E}}^{[12]}$ in the $3\alpha$ model (See, e.g., Ref. \cite{Ya08}).
Thus, the reduced electric multipole transition strength distribution is obtained as 
\begin{eqnarray}
\frac{dB(\mathrm{E}\lambda:J_i \to J_f)}{dE} &=&  
\frac{4 e^2}{2 J_i+1}
\sum_{M_i, M_f, \mu} \int d\bm{q}^\prime d\bm{p}^\prime 
\left\vert \langle \Psi^{(-)}_{\bm{q}^\prime,\bm{p}^\prime} (J_f M_f, E^\prime)\vert \Omega_{\lambda \mu} \vert  \Psi_b (J_i M_i) \rangle  \right\vert^2
\cr
&& \times \delta\left(E- E^\prime \right).
\label{eq:SE_psi}
\end{eqnarray} 
Here $\vert \Psi^{(-)}_{\bm{q},\bm{p}}(J_f M_f,E) \rangle$ is a partial wave component of  the final $3\alpha$ continuum state
of  angular momentum $J_f$ and its third component $M_f$,  and  $\Omega_{\lambda \mu}$ is an isoscalar multipole operator as conventionally defined by 
\begin{equation}
\Omega_{\lambda \mu}  = 
\left\{
\begin{array}{cc}
 \sum_{i=1,3}  r_i^2  \quad\quad & (\mathrm{for}~ \lambda=0)
\\
\sum_{i=1,3} r_i^3 Y_{1}^{\mu}(\hat{\bm{r}}_i)  \quad\quad & (\mathrm{for}~  \lambda=1)
\\
\sum_{i=1,3} r_i^{\lambda}  Y_{\lambda}^{\mu}(\hat{\bm{r}}_i)   \quad\quad & (\mathrm{otherwise}). 
\end{array}
\right.
\label{eq:12C_gs-E2-3a-2}
\end{equation} 

A peak in the transition strength distribution as a function of $E$ may correspond to a resonance state, whose line shape  is fitted by a resonance formula:
\begin{equation}
\frac{dB}{dE}=\frac{B}{\pi}\frac{\Gamma_{3\alpha}/2}{(E-E_r)^2 + (\Gamma_{3\alpha}/2)^2},
\label{eq:dBdE-BW}
\end{equation}
from which the resonance energy $E_r$, the $3\alpha$-decay width $\Gamma_{3\alpha}$, and the strength $B$ are evaluated.
Note that the integration of Eq. (\ref{eq:dBdE-BW}) around $E_r$ agrees well  with the strength value $B$ if the width $\Gamma_{3\alpha}$ is small enough. 

%%%%%%%%%%%%%%%%%%%%%%%%%%%%%%%%
\section{Calculations of multipole strength distributions}
\label{sec:strength_distributions}
%%%%%%%%%%%%%%%%%%%%%%%%%%%%%%%

%%%%%%%%%%%%%
\subsection{Interaction models}
%%%%%%%%%%%%%

For the $2\alpha$ interaction potential,  I use  the form of the Ali-Bodmer (AB) model \cite{Al66}, which is 
a state-dependent two-range Gaussian nuclear potential along  with the point $\alpha$-$\alpha$ Coulomb potential, 
\begin{equation}
V(x) = \left( V_\mathrm{R}^{(0)} \hat{P}_{L=0} + V_\mathrm{R}^{(2)} \hat{P}_{L=2} \right) e^{-(x/a_\mathrm{R})^2}  
 + V_\mathrm{A} e^{-(x/a_\mathrm{A})^2} + \frac{4e^2}{x}.
\label{eq:aa-pot}
\end{equation} 
Here, $x$ is the distance between two $\alpha$-particles and $\hat{P}_{L}$ is the projection operator on the $2\alpha$ state of  angular momentum $L$. 
In the present work, two different versions of the AB model are used:
 one presented in Ref. \cite{Fe96} (slightly modified version of model A in Ref. \cite{Al66}), AB(A'),  and the model D given in Ref.  \cite{Al66}, AB(D). 
Values of the range- and strength parameters of these models are shown in Table \ref{tab:aa-pot-parameters} together with  calculated values of $\alpha$-$\alpha$ observables in comparing with the empirical values \cite{Ti04}. 
It is noted that the attractive tail in the nuclear part of AB(D)  has a shorter range than that of  AB(A'), and  that  AB(D) gives a better fit to the experimental phase shifts than  AB(A') as suggested in Ref. \cite{Al66}. 

%%%%%%%%%%%%%%%%%%%%%%%%%%%%%%%%%%%%%%%
%%% Table II: alpha-alpha potential parameters 
%%%%%%%%%%%%%%%%%%%%%%%%%%%%%%%%%%%%%%%
\begin{table}[t]
\caption{\label{tab:aa-pot-parameters}
Parameters of the $2\alpha$ potential, Eq. (\ref{eq:aa-pot}),  
for  the AB(A') \cite{Fe96} and the AB(D) \cite{Al66} models, and calculated values of the resonance energy $E_{r,2\alpha}$ and the width $\Gamma_{2\alpha}$ for $L=0$ and $L=2$ states.  
Experimental values are taken from Ref. \cite{Ti04}. 
}
\begin{ruledtabular}
\begin{tabular}{cccc} 
 & AB(A') & AB(D)  & Exp. \\
\hline
$a_\mathrm{R}$ [fm]             & 1.53   & $1/0.70 ~(\approx1.4)$  &\\
$V_\mathrm{R}^{(0)}$ [MeV] & 125.0 & 500.0 &  \\
$V_\mathrm{R}^{(2)}$ [MeV]  &  20.0  &  320.0\\
$a_\mathrm{A}$ [fm]               & 2.85& $1/0.475~ (\approx2.11)$  \\
$V_\mathrm{A}$ [MeV]          &  -30.18 & -130.0 \\ 
$E_{r,2\alpha(L=0)}$ [keV]            & 93.4     &  95.1    &  91.8  \\
$\Gamma_{2\alpha(L=0)}$ [eV]     &   8.59   &  8.32  &   $5.57\pm0.25$ \\
$E_{r,2\alpha(L=2)}$ [MeV]            & 3.47     &  3.38    &  $3.12\pm0.01$  \\
$\Gamma_{2\alpha(L=2)}$ [MeV]     &   3.81   &  2.29  &   $1.513\pm0.015$ \\
\end{tabular}
\end{ruledtabular}
\end{table}
%%%%%%%%%%%%%%%%%%%%%%%%%%%%%%%%%%%%%%%%%%%%%%%%%%%%%%%%%%%%%%%%%%%%

In obtaining solutions of Eq.  (\ref{eq:Psi_def}),  $3\alpha$ partial wave states with the angular momentum of the $2\alpha$ subsystem with 0, 2, and 4 are taken into account.

Calculated $3\alpha(0^+)$ energies for  the AB(A')-$2\alpha$P and AB(D)-$2\alpha$P are  $-0.832$ MeV and  $-1.451$ MeV, respectively, which are less bound compared to the empirical value (See Table  \ref{tab:12C-levels}).
Like this, in general, the use of only $2\alpha$P  is not sufficient in reproducing ${}^{12}$C observables such as the binding energies and the resonance energies.
In order to reproduce these observables, I introduce a three-body potential, which depends on the total angular momentum of the $3\alpha$ system, 
\begin{equation}
V_{3\alpha} = \sum_{J,\pi}   \hat{P}_{3\alpha,J^\pi}
W_3^{(J^\pi)} 
\exp\left(-\frac{r_{12}^2  + r_{23}^ 2+  r_{31}^2 }{a_3^2} \right), 
\label{eq:3bp}
\end{equation}
where $\hat{P}_{3\alpha,J^\pi}$ is a projection operator on the $3\alpha$ state with the total angular momentum $J$ and parity $\pi$, and $r_{ij}$ is the distance between particles $i$ and $j$.
In this work, taking the convention of Ref. \cite{Fe96},  $a_3=\sqrt{\frac{3}{3.97}}b_3$, 
I use four different values for $b_3$:  4.6 fm, 3.9 fm, 3.45 fm, and 3.0 fm, which correspond to $a_3= 4.0$ fm, 3.4 fm, 3.0 fm, and 2.6 fm, respectively.
When combined with AB(A') [AB(D)], these $2\alpha$P + $3\alpha$P models will be designated as  A1, A2, A3, and A4 [D1, D2, D3, and D4]  models, respectively. 

In the present work, I will consider  $3\alpha$ states of  $J^\pi=0^+,1^-,2^+$, and $3^-$.
For each of $3\alpha$ state $J^\pi$, the strength parameter $W_3^{(J^\pi)}$ is determined to reproduce the following observables: 
the resonance energy of the Hoyle state ${}^{12}$C$(0_2^+)$  for $J=0$ state, the binding energy of first excited ${}^{12}$C$(2_1^+)$ state for $J=2$, 
the lowest resonance energies of ${}^{12}$C$(1_1^-)$ state for $J=1$  and  ${}^{12}$C$(3_1^-)$ state for $J=3$.
Thus determined parameters are summarized in  Table \ref{tab:W3_data}, and numerical results with these sets of potentials will be presented in the following subsections .

%%%%%%%%%%%%%%%%%%%%%%%%%%%%%%%%%%%%%%%
%%% Table III: $3\alpha$P strength parameters
%%%%%%%%%%%%%%%%%%%%%%%%%%%%%%%%%%%%%%%
%longtable
\begin{table}[t]
\caption{\label{tab:W3_data} 
The range- and strength parameters of the $3\alpha$ potential.
}
\begin{ruledtabular}
\begin{tabular}{cccccc}
   &  $a_3$ & $W_3^{(0^{+})}$  &  $W_3^{(1^{-})}$ &  $W_3^{(2^{+})}$  &  $W_3^{(3^{-})}$  \\   
Model &  fm  &  MeV &  MeV &  MeV &  MeV \\  
\hline 
AB(A') + $3\alpha$P \\
A1 &  4.0  & -51.99 & -11.4 & -33.3  & -5.10\\ 
A2 & 3.4  & -96.0 &  -26.91 & -56.3 & -11.13  \\ 
A3 & 3.0 & -150.84  & -54.6  & -87.8 & -22.0\\  
A4 & 2.6   &-247.26 & -129  & -151.65 & -52.8  \\ 
AB(D) + $3\alpha$P \\
D1  & 4.0   & -48.54 & -15.0 & -25.5   & 6.03 \\ 
D2  &  3.4 & -92.85 & -37.0 & -46.0  & 12.69 \\
D3  & 3.0 & -156.9  & -79.0 &-78.3 & 25.0 \\  
D4  & 2.6 &-303.66 &-193 & -158.1  &  63.6\\  
\end{tabular}
\end{ruledtabular}
\end{table}
%%%%%%%%%%%%%%%%%%%%%%%%%%%%%%%%%%%%%%%%%%%%%%%%%%%%%%%%%%%

%%%%%%%%%%%%
\subsection{ $3\alpha (0^+)$ states}
%%%%%%%%%%%

%%%%%%%%%%%%%%%%%%%%%%%
\subsubsection{${}^{12}\rm{C}(0_1^+)$ state}
%%%%%%%%%%%%%%%%%%%%%%%%

Calculated values of the $3\alpha(0^+)$ binding energy [$\vert E(0_1^+) \vert $]  are shown in Table \ref{tab:12C_gs}.
The binding energies for the AB(A') + $3\alpha$P models are rather small compared to the experimental value, while those for the AB(D) + $3\alpha$P models fairly agree with the experimental value. 
It is noted that the calculated binding energy is not monotonically changing as a function of the $3\alpha$P range parameter $a_3$ in Eq. (\ref{eq:3bp}), and there is a value of $a_3$ that gives the largest $3\alpha$ binding energy (or the minimum energy).
From Table \ref{tab:12C_gs}, it is suggested that the use of the AB(A')-$2\alpha$P together with an one-range Gaussian $3\alpha$P, Eq. (\ref{eq:3bp}), is not able to reproduce the energies of  ${}^{12}\rm{C}(0_1^+)$ and ${}^{12}\rm{C}(0_2^+)$ states at the same time.

The charge radius of ${}^{12}$C in the $3\alpha$ model is calculated by
\begin{equation}
r_{\mathrm{ch}} = \sqrt{ r_{\alpha,\mathrm{ch}}^2 + r_{\mathrm{rms}}^2 },
\label{eq:r_ch}
\end{equation}
where $r_{\mathrm{rms}}$ is the root mean square (r.m.s.) radius of the $3\alpha$-system with the wave function $\Psi_b$,
\begin{eqnarray}
 r_{\mathrm{rms}} &=& \sqrt{\frac13 \sum_{i=1,3} \langle \Psi_b \vert r_i^2 \vert \Psi_b \rangle }
\cr
&=&  \sqrt{\frac16 \langle \Psi_b \vert x^2 \vert \Psi_b \rangle +\frac29 \langle \Psi_b \vert y^2 \vert \Psi_b \rangle },
\end{eqnarray}
and $r_{\alpha,\mathrm{ch}}$ is the charge radius of the $\alpha$-particle.
An experimental value for  $r_{\mathrm{rms}}$ is obtained as $r_{\mathrm{rms}}=1.8823(36)$ fm using the experimental values: $r_{\mathrm{ch}}=2.4829(19)$ fm \cite{Ke17,Ru84} and  $r_{\alpha,\mathrm{ch}}= 1.6755(28)$ fm \cite{An13}, in Eq. (\ref{eq:r_ch}). 
Calculated values of $r_{\mathrm{rms}}$ are also shown in Table \ref{tab:12C_gs}.
It is interesting that the calculated radii almost scale to the $3\alpha$P range parameter, which gives $a_3=3.6$ fm to reproduce the experimental  $r_{\mathrm{rms}}$.

%%%%%%%%%%%%%%%%%%%%%%%%%%%%%%%%%%%%%%%
%%% Table IV: 12C_gs
%%%%%%%%%%%%%%%%%%%%%%%%%%%%%%%%%%%%%%%
\begin{table}[t]
\caption{\label{tab:12C_gs}
Calculated values of the binding energy and the r.m.s. radius for ${}^{12}\mathrm{C}(0_1^+)$ state.  }
\begin{ruledtabular}
\begin{tabular}{ccc}
            & Binding energy  &  $r_{\mathrm{rms}}$ \\ 
 Model & MeV  &  fm \\
\hline
Exp. & 7.2746  & 1.8823(36) \\
AB(A') + $3\alpha$P \\
A1 & 6.337 &  1.96 \\  
A2 & 6.725 &  1.84 \\  
A3 & 6.527 &  1.77 \\   
A4 & 5.554 & 1.69 \\  
AB(D) + $3\alpha$P \\
D1 & 7.281 & 1.93 \\   
D2 & 7.789 & 1.86 \\   
D3 & 7.759 & 1.80 \\ 
D4 & 7.584 & 1.71 \\  
\end{tabular}
\end{ruledtabular}
\end{table}
%%%%%%%%%%%%%%%%%%%%%%%%%%%%%%%%%%%%%%%

%%%%%%%%%
\subsubsection{Hoyle state}
%%%%%%%%

Calculated E0 transition strength distribution (TSD) from ${}^{12}\mathrm{C}(0_1^+)$  to $3\alpha(0^+)$ continuum states has a sharp peak as a function of $E$ just above the $3\alpha$ breakup threshold, which corresponds to the Hoyle state. 
%<><><
Its line shape  is fitted by Eq. (\ref{eq:dBdE-BW}), from which the resonance energy $E_r(0_2^+)$, the $3\alpha$-decay width $\Gamma_{3\alpha}(0_2^+)$, and the E0 strength  $B(\mathrm{E}0:0_1^+ \to 0_2^+)$ are evaluated. 
These results together with the peak values of the E0-TSD are shown in Table \ref{tab:Hoyle-state}. 

%%%%%%%%%%%%%%%%%%%%%%%%%%%%%%%%%%%%%%%
%%% Table V: Hoyle state
%%%%%%%%%%%%%%%%%%%%%%%%%%%%%%%%%%%%%%%
\begin{table}[t]
\caption{\label{tab:Hoyle-state}
Calculated values of the resonance energy, peak value of E0-TSD, $3\alpha$-decay width, and E0 strength for for the Hoyle state.  }
\begin{ruledtabular}
\begin{tabular}{ccccc}
 &  $E_r(0_2^+)$ &$\frac{dB}{dE}\vert_{E=E_r}$ & $\Gamma_{3\alpha}(0_2^+)$  & $B(\mathrm{E0}:0_1^+\to 0_2^+)$ \\
 Model &   MeV  &  $10^6 \frac{e^2\mathrm{fm}^4}{\mathrm{MeV}}$ & eV  &  $e^2$fm$^4$  \\ 
\hline
Exp. &  0.3795 & &  9.3(9)\footnote{Ref. \cite{Ke17}.}  &  29.9(1.0)\footnote{Ref. \cite{Ch10}.}  \\ 
AB(A') + $3\alpha$P \\
A1 & 0.379469 & 3.0 &10.1& 48.1 \\ \ 
A2 & 0.379302 & 2.6 &10.0& 40.7\\  
A3 & 0.379166 & 2.6 &9.9 & 40.1 \\      
A4 & 0.379614 & 3.2 & 9.9 & 50.3 \\    
AB(D) + $3\alpha$P \\
D1 &  0.379222 & 3.2 & 6.8 & 34.8 \\   
D2 &  0.379177 & 2.8 &6.2 & 28.0 \\  
D3 &  0.379278 & 2.6 & 6.0 & 24.9 \\    
D4 & 0.379417 & 1.8 & 6.0 & 17.1 \\    
\end{tabular}
\end{ruledtabular}
%Data from refs.
\end{table}
%%%%%%%%%%%%%%%%%%%%%%%%%%%%%%%%%%%%%%%

It is noted that  $\Gamma_{3\alpha}(0_2^+)$ is not directly measured, but  was obtained from the combination of  $B(\mathrm{E}0:0_1^+ \to 0_2^+)$ and  the E0 electron-positron pair branching ratio of the Hoyle state, $\Gamma_{\pi}(\mathrm{E0})/\Gamma(0_2^+)$.
Here, $\Gamma_{\pi}(\mathrm{E}0)$ is the E0 pair decay width of the transition process ${}^{12}\mathrm{C}(0_2^+) \to {}^{12}\mathrm{C}(0_1^+) + e^+ + e^-$,  and  $\Gamma(0_2^+)$ is the total decay width of the Hoyle state, which is essentially equal to $\Gamma_{3\alpha}(0_2^+)$. 
The pair decay width $\Gamma_{\pi}(\mathrm{E}0)$ is related to $B(\mathrm{E}0:0_1^+ \to 0_2^+)$ as given in Refs. \cite{Wi69,Ch10}.
A global fit to experimental data of the inelastic electron scattering ${}^{12}\mathrm{C}(e,e^\prime)$ with  the momentum transfer up to $3.1~\mathrm{fm}^{-1}$ \cite{Ch10} gives
$B(\mathrm{E}0:0_1^+ \to 0_2^+)=29.9(1.0) ~e^2\mathrm{fm}^4$, which gives  $\Gamma_{\pi}(\mathrm{E0})=62.3(2.0) ~\mu\mathrm{eV}$. 
The adapted value of the branching ratio in the compilation \cite{Ke17},  $\Gamma_{\pi}(\mathrm{E0})/\Gamma(0_2^+)=6.7(6)\times10^{-6}$,  gives $\Gamma_{3\alpha}(0_2^+)=9.3(9)$ eV in Table \ref{tab:Hoyle-state}. 

Calculated values of $\Gamma_{3\alpha}(0_2^+)$ do not depend so much on the choice of $3\alpha$P, but depend on the choice of the $2\alpha$P. 
While those of the AB(A') + $3\alpha$P models are consistent with the experimental value above, those of the AB(D) + $3\alpha$P models are about 40\% smaller than this.

Recently, a slightly larger value than the previous E0 branching ratio, $\Gamma_{\pi}(\mathrm{E0})/\Gamma(0_2^+)=8.2(5)\times10^{-6}$, has been obtained  from an analysis of  ${}^{12}\mathrm{C}(p,p^\prime)$ reaction \cite{Er20}, which gives  $\Gamma_{3\alpha}(0_2^+)=7.6(5)$ eV.
This value is closer to the AB(D) + $3\alpha$P calculations than the AB(A') + $3\alpha$P calculations. 

%%%%%%%%%
\subsubsection{Higher $3\alpha (0^+)$  resonance states}
\begin{figure}[tb]
\includegraphics[width=0.9\columnwidth]{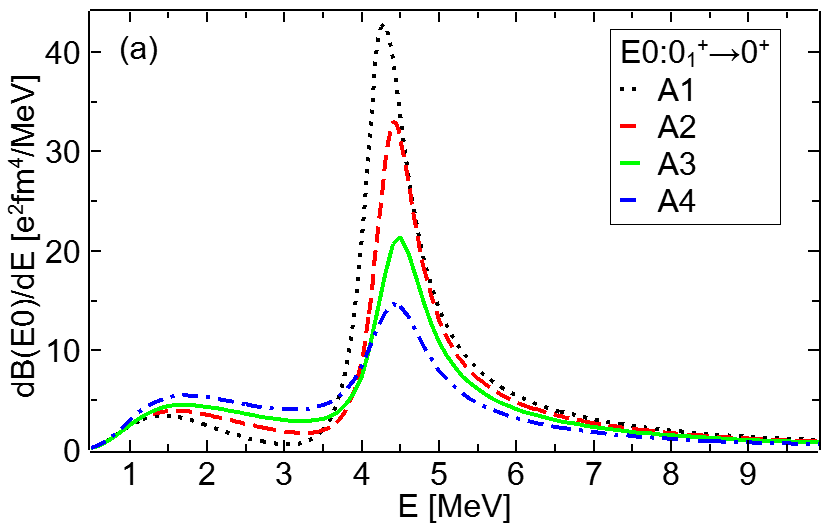}
\includegraphics[width=0.9\columnwidth]{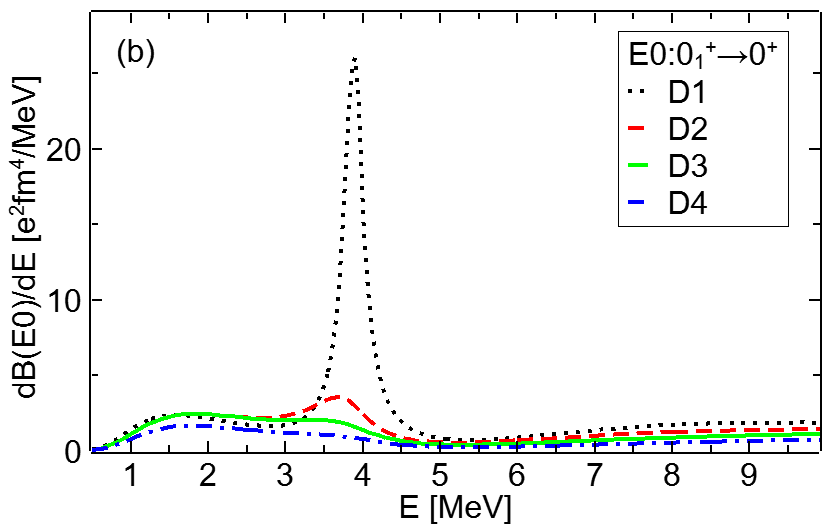}
\caption{(Color online) 
The E0-TSDs for the transition from ${}^{12}$C${(0_1^+)}$ to $3\alpha(0^+)$ states. 
Dotted (black), dashed (red), solid (green), and dot-dashed (blue) curves denote calculations with A1 [D1], A2 [D2], A3 [D3], and A4 [D4] models, respectively in (a) [(b)]. 
Note that peaks of the Hoyle state are not displayed in these figures, but their peak values are shown in Table \ref{tab:Hoyle-state}.
\label{fig:be0-0-0}
}
\end{figure}

%%%%%%%%
The E0-TSDs for the energy range of $0.5~ \mathrm{MeV} < E < 10~ \mathrm{MeV}$ calculated for the AB(A') + $3\alpha$P and  AB(D) + $3\alpha$P models are shown in Fig. \ref{fig:be0-0-0} (a) and (b), respectively.
At this energy range, there are two peaks for all calculations, which will be denoted as $0_3^+$ and $0_4^+$. 
While the $0_3^+$ peaks for all models are located around $E=1.5$ MeV, the energies of the $0_4^+$ peaks are model dependent and located in the range of 3.5 to 4.5 MeV. 
Also, it is remarkable that  the strength of the $0_4^+$ state is strongly model dependent, and it has a tendency that  the strength  becomes larger as interaction range is larger.

%%%%%%%%%%%%%%%%%%%%%%%%%%%%%%
%----- FIGURE 1 -------------------------------------------------------
%%%%%%%%%%%%%%%%%%%%%%%%%%%%%%
%---------------------------------------------------------------------

There have been some debates on the nature of the $0_3^+$ peak. 
The energy spectrum for a reaction to produce a sharp resonance state just above the threshold energy, e.g., ${}^{9}\mathrm{Be}(p,d){}^{8}\mathrm{Be}$ reaction,  sometimes reveals an additional  small and broad peak at an energy above the resonance peak  \cite{Be71,Be81}, which is called as Ghost Anomaly.
In a R-matrix theory of the resonance formula \cite{Ba62},  this peak is considered to be caused as a result of  rather rapid energy-dependence of a resonance decay width parameter arising from  the Coulomb penetration factor.

In recent analysis of the inclusive excitation-energy spectra in  ${}^{12}\mathrm{C}(\alpha,\alpha^\prime){}^{12}\mathrm{C}$ and ${}^{14}\mathrm{C}(p,t){}^{12}\mathrm{C}$ reactions in terms of a multi-level, multi-channel R-matrix theory \cite{Li22a,Li22b}, 
the $0_3^+$ peak (denoted as $0_\Delta^+$ in the references) turns to be an additional resonance state because the Ghost peak associated with the Hoyle state does not have enough strength to explain the experimental data. 

\begin{figure}[tb]
\includegraphics[width=1.0\columnwidth]{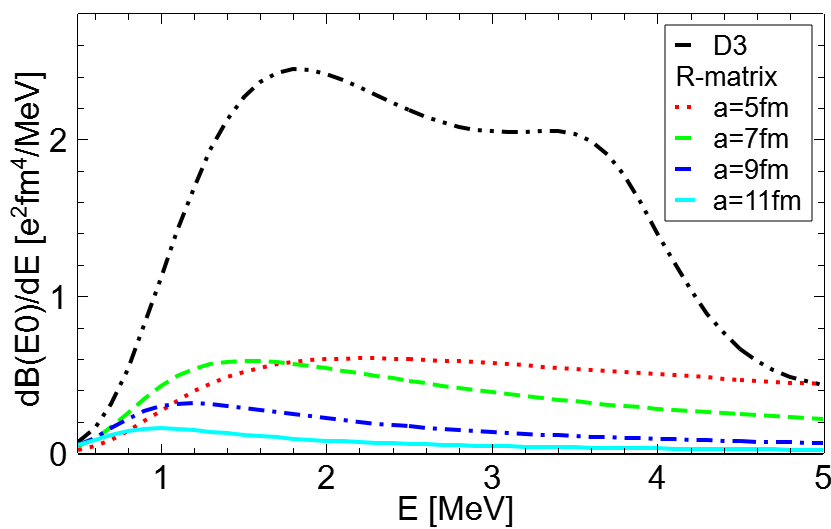}
%,angle=-90
\caption{(Color online) 
The E0-TSDs for the transition from ${}^{12}$C${(0_1^+)}$ to $3\alpha(0^+)$ states.
Dot-dot-dashed (black) curve denotes the calculations of the D3 model.
Dotted (red), dashed (green), dot-dashed (blue), and solid (cyan) curves denote calculations by the R-matrix formula with the channel radius parameters $a=$ 5 fm, 7 fm, 9 fm, and 11 fm, respectively.
\label{fig:be0-D3-R-matrix}
}
\end{figure}

To examine the $0_3^+$ peak in the present $3\alpha$ model, the Hoyle state peak for the D3 model is fitted by R-matrix parameterization as in Ref. \cite{Li22a}.
In Fig. \ref{fig:be0-D3-R-matrix}, thus calculated E0-TSDs in R-matrix with taking several values of the channel radius parameter $a$, are compared with the D3 result.
As shown in Fig. \ref{fig:be0-D3-R-matrix}, the R-matrix calculations reveal the Ghost peak, but their strengths are quite small compared to the D3 calculation, which is not enough to reproduce the experimental data as in Ref. \cite{Li22a}.
As the channel radius $a$ increases, the energy  of the Ghost peak is shifted to lower energy and  the strength becomes weak. 
Note that the channel radius $a=11$ fm was used in Ref. \cite{Li22a}. 
Thus, a large part of the $0_3^+$ peak may be caused by a resonance pole with a minor contribution of the Ghost peak.

%%%%%%%%%%%%%%%%%%%%%%%%%%%%%%%%%%
%----- FIGURE 2 -------------------------------------------------------
%%%%%%%%%%%%%%%%%%%%%%%%%%%%%%%%%%
%---------------------------------------------------------------------
%%%%%%%%%%%%%%%%%%%%%%%%%%%%%%%

%%%%%%%%%%%%%
\subsection{$3\alpha(2^+)$ states}
%%%%%%%%%%%%%

%%%%%%%%%%%%%%%%%%%%%%%
\subsubsection{${}^{12}\rm{C}(2_1^+)$ state}
%%%%%%%%%%%%%%%%%%%%%%%%

As mentioned previously, the strength parameter of the $3\alpha$P for $3\alpha(2^{+})$ state  is determined to reproduce the energy of the first excited state ${}^{12}\rm{C}(2_1^+)$.
In Table \ref{tab:C12_2_1_results}, calculated energies, the reduced  E2 transition strength between the ground state and the first excited state, $B(\mathrm{E2}:2_1^+ \to 0_1^+)=B(\mathrm{E2}:0_1^+ \to 2_1^+)/5$,  and the quadrupole moment $Q_\mathrm{S}(2_1^+) = 0.75793 \langle 2_1^+ \vert \vert \mathrm{E}2 \vert\vert 2_1^+ \rangle $ \cite{Sa23},  are shown. 
%, which is related to the reduced E2 matrix element of the $2_1^+$ state by $Q_\mathrm{S}(2_1^+)
In the recent measurement in an electron scattering \cite{DA20}, the value of $7.63(19) ~e^2  \mathrm{fm}^4$ is obtained for $B(\mathrm{E2}:2_1^+ \to 0_1^+)$, which is consistent with other electron scattering results reported in the compilation \cite{Pr16}:   $[7.94(66), 7.72(74), 8.12(82), 9.4(2.0) ]~e^2  \mathrm{fm}^4$, and also with those obtained from $(\alpha,\alpha^\prime)$ reactions:  $7.68(1.00) ~ e^2  \mathrm{fm}^4$ \cite{Jo03} and $7.4(2) ~ e^2  \mathrm{fm}^4$ \cite{It11}. 

The calculated values of  $B(\mathrm{E2}:2_1^+ \to 0_1^+)$  correlate well with the $3\alpha$P range parameter $a_3$ in Eq. (\ref{eq:3bp}),  and the experimental value of \cite{DA20} gives the range parameter $a_3 =  3.7 $ fm for the AB(A') + $3\alpha$P calculations and $a_3 =  3.1$ fm for  the AB(D) + $3\alpha$P calculations, which supports the A1, A2, D2, and D3 models.

Also in Ref. \cite{DA20}, using a correlation between calculated values of $B(\mathrm{E2}:2_1^+ \to 0_1^+)$ and those of the quadrupole moment $Q_\mathrm{S}(2_1^+)$ obtained with {\it ab initio} calculations based on the no-core shell model \cite{Ge17},  $Q_\mathrm{S}(2_1^+)=5.97(30) ~ e \mathrm{fm}^2$ was extracted.

For the present $3\alpha$ model calculations,  the  correlation between  $B(\mathrm{E2}:2_1^+ \to 0_1^+)$  and  $Q_\mathrm{S}(2_1^+)$  is also observed for each of  the $2\alpha$P.
The use of the AB(A') + $3\alpha$P models results  $Q_\mathrm{S}(2_1^+)=6.41(46) ~ e \mathrm{fm}^2$ while that of AB(D) + $3\alpha$P models does $Q_\mathrm{S}(2_1^+)=6.09(18) ~ e  \mathrm{fm}^2$.
These values agree with the one obtained in Ref. \cite{DA20}, which suggests the correlation between  $B(\mathrm{E2}:2_1^+ \to 0_1^+)$  and $Q_\mathrm{S}(2_1^+)$ is less nuclear model dependent.

The experimental value of  $Q_\mathrm{S}(2_1^+)$ has been measured by the Coulomb-excitation of ${}^{12}$C by a heavy ion target.
The theoretical values above obtained for $Q_\mathrm{S}(2_1^+)$ are consistent with experimental value of Ref.  \cite{Ve83},
$Q_\mathrm{S}(2_1^+)=+6(3) ~e \mathrm{fm}^2$, and  that of Ref.  \cite{Ra18},   
$Q_\mathrm{S}(2_1^+)=+7.1(2.5) ~e \mathrm{fm}^2$, but not consistent with the recent value, 
$Q_\mathrm{S}(2_1^+)=+9.5(1.8) ~e \mathrm{fm}^2$, obtained in Ref. \cite{Sa23}.

%%%%%%%%%%%%%%%%%%%%%%
%%% Table VI:  2^+
%%%%%%%%%%%%%%%%%%%%
\begin{table}[t]
\caption{
Calculated values of the binding energy, E2 transition strength, and quadrupole moment of ${}^{12}\rm{C}(2_1^+)$.
\label{tab:C12_2_1_results} 
}
\begin{ruledtabular}
\begin{tabular}{cccc}
       & Binding energy & $B(\mathrm{E2}:2_1^+ \to 0_1^+)$   & $Q_\mathrm{S}(2_1^+)$   \\
Model     &   MeV & $e^2\mathrm{fm}^4$  & $e\mathrm{fm}^2$ \\
\hline
Exp. & 2.8348\footnote{Ref. \cite{Ke17}.} & 7.63(19)\footnote{Ref. \cite{DA20}.}   &  6(3)\footnote{Ref. \cite{Ke17}.} \\
AB(A') + $3\alpha$P \\
A1 &   2.848 & 8.68  &  6.65\\
A2 & 2.841 & 6.86 & 6.29 \\
A3 & 2.836 & 5.74 & 5.95 \\
A4 & 2.840 & 4.66 & 5.43 \\
AB(D) + $3\alpha$P \\
D1 &  2.851 & 9.43 &   6.46 \\ 
D2 & 2.830& 8.20 & 6.25 \\
D3 & 2.832 &7.32 & 6.05 \\ 
D4 & 2.837 & 6.15 & 5.78 \\ 
%D5 & 2.822 & 5.29 & 5.48 
\end{tabular}
\end{ruledtabular}
\end{table}
%%%%%%%%%%%%%%%%%%%%%

%%%%%%%%%%%%%%%%%%%%%%%%% 
\subsubsection{$3\alpha(2^+)$ continuum states}
\begin{figure}[tb]
\includegraphics[width=0.6\columnwidth]{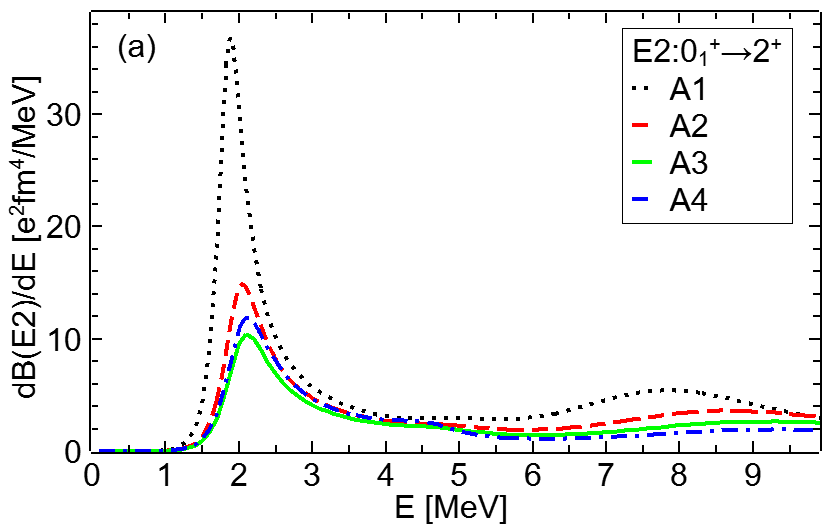}
\includegraphics[width=0.6\columnwidth]{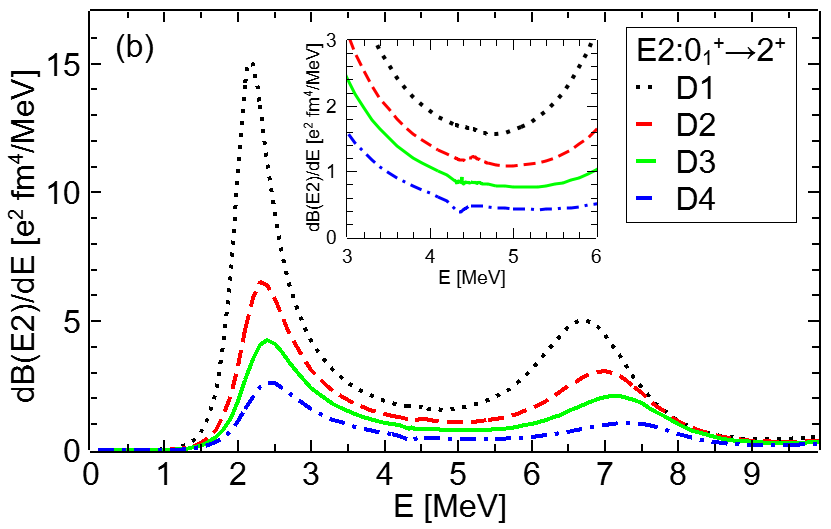}
\includegraphics[width=0.6\columnwidth]{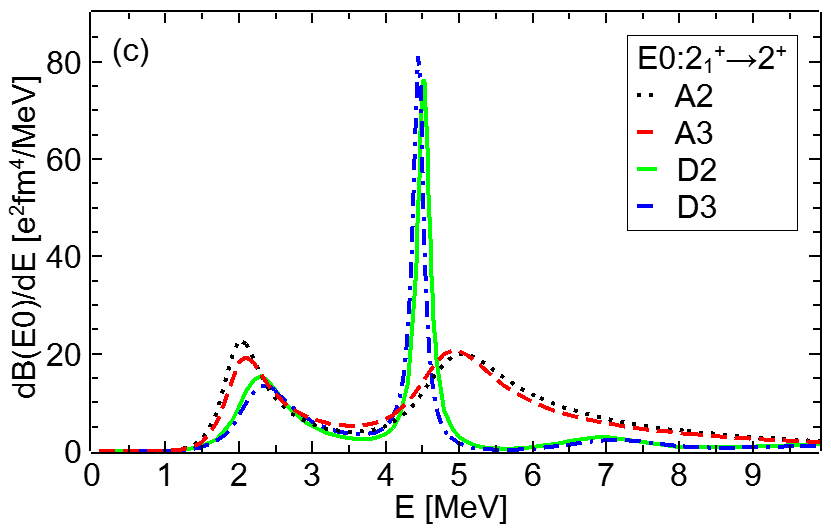}
%,angle=-90
\caption{(Color online) 
(a) and (b): The E2-TSDs  for the transition from ${}^{12}$C${(0_1^+)}$ to $3\alpha (2^+)$ states.
% for (a) the AB(A')-$3\alpha$P models, (b) the  AB(D)-$3\alpha$P models, and 
(c) The E0-TSDs for the transition from ${}^{12}$C${(2_1^+)}$ to $3\alpha(2^+)$ states. 
In (a) [(b)], dotted (black), dashed (red), solid (green), and dot-dashed (blue) curves denote the calculations of the  A1 [D1], A2 [D2], A3 [D3], and A4 [D4] models, respectively.
Inset in (b) is the magnification around $E=4.5$ MeV.
In (c), dotted (black), dashed (red), solid (green), and dot-dashed (blue) curves denote calculations with the A2, A3, D2, and D3 models, respectively.
\label{fig:db-de-2}
}
\end{figure}

%%%%%%%%%%%%%%%%%%%%%%%%%
%
Calculated E2-TSDs for the transition from the ${}^{12}$C${(0_1^+)}$ state to $3\alpha(2^+)$ continuum states are shown in Fig. \ref{fig:db-de-2} (a) for the  AB(A') + $3\alpha$P models and in Fig. \ref{fig:db-de-2} (b) for the AB(D) + $3\alpha$P models as functions of the energy $E$. 
A common feature of all the calculated E2-TSDs is the existence of peaks at $E \approx 2$ MeV and at  $E \approx 7$ MeV, although the latter peaks for the AB(A') + $3\alpha$P models are rather indistinct.
In addition to these peaks, a tiny structure  can be seen around  $E=4.5$ MeV in the E2-TSDs as demonstrated in the inset of Fig. \ref{fig:db-de-2} (b). 
The existence of $3\alpha(2^+)$ state at this energy region has been reported previously in analises of 
 ${}^{11}\mathrm{B}({}^{3}\mathrm{He},d){}^{12}\mathrm{C}$ reaction \cite{Re71},
${}^{12}$B and ${}^{12}$N $\beta$-decay data \cite{Hy10}, and $(\alpha,\alpha^\prime)$ reaction \cite{Jo03}.
This  $2^+$ state was once adapted as $(E_r, \Gamma_{3\alpha})=$(3.89 MeV, 0.43 MeV)  in the compilations \cite{Aj85,Aj90}, but disappears in the most recent compilation \cite{Ke17} after the experimental reports to show negative evidence for it \cite{Zi11,Sm12}.  
Theoretically, the existence of $3\alpha(2^+)$ state around $E=4$ MeV has been shown in semi-microscopic $3\alpha$ cluster model calculations  \cite{Ku07,Oh13}  and a  $3\alpha$ model calculation with the hyperspherical adiabatic formalism  \cite{Ga15}.

Here, I examined a monopole transition to $3\alpha(2^+)$ continuum states from the ${}^{12}$C${(2_1^+)}$ state instead of the ${}^{12}$C${(0_1^+)}$ state. 
In  Fig. \ref{fig:db-de-2} (c), the E0-TSDs for this transition for the A2, A3, D2, and D3 models are shown.
These functions show clear peaks around $E \approx 4.5$ MeV, which indicates the existence of a resonance state.
Meanwhile, as demonstrated in Figs. \ref{fig:db-de-2} (a) and (b), the E2 transition strength of this state from the ${}^{12}$C${(0_1^+)}$ state is small, and this might be the reason why the existence of this state has not been clearly established.
 
Hereafter, the peaks at $E \approx 2$ MeV, $E \approx 4.5$ MeV, and $E \approx 7$ MeV will be designated as $2_2^+$, $2_3^+$, and $2_4^+$ states, respectively. 
The resonance parameters, $E_r$ and $\Gamma_{3\alpha}$, which are obtained by analyzing adequate peaks of the TSDs in Fig. \ref{fig:db-de-2} with Eq. (\ref{eq:dBdE-BW}) are shown in Table \ref{tab:C12_2_energies}. 

As mentioned in the introduction, the existence of the  $2_2^+$ state in ${}^{12}$C has been reported in some experimental works
and the experimental values of $\left( E_r, \Gamma_{3\alpha} \right)$ in MeV are 
$\left[2.3(1), 0.6(1)\right]$  in $(p,p^\prime)$ data \cite{Fr09,Zi11},   % 9.6(1) 0.6(1)
$\left[2.57(6), 1.01(15)\right]$ in $(\alpha,\alpha^\prime)$ data \cite{It11},  %9,84(6), 1.01(15)
$\left[2.48(15),  0.750(150)\right]$ in combined analysis  \cite{Fr12} of the  data in Refs. \cite{Fr09,It11},  % 9.75(15)   0.75(15)
$\left[2.76(11), 0.800 (130)\right]$ in $(\gamma, \alpha)$  \cite{Zi13}, % 10.03(11)  0.80(13)
and   
$\left[2.751(50), 1.60(13)\right]$ in reanalysis of the data  \cite{Zi13}, which is cited in Ref. \cite{Fr18}. % 10.025(50) 1.60(13)
The calculated  values of the resonance energy and the $3\alpha$-decay width from by the present $3\alpha$ calculations in Table \ref{tab:C12_2_energies}  are rather model independent and are consistent with these experimental values.

The E2 strength of the $2_2^+$ state $B(\mathrm{E2}:0_1^+ \to 2_2^+)$ in Table \ref{tab:C12_2_energies} is evaluated by integrating the E2-TSD for $E < 3.5$ MeV, which indicates that this strength largely depends on the range parameters of $2\alpha$P and $3\alpha$P.
It is noted that theoretical values by microscopic calculations also show a large model dependence: $2~e^2\mathrm{fm}^4$ \cite{Ka07,Ko11} and $10(5)~e^2\mathrm{fm}^4$  \cite{Ep12}.

Experimental results for  $B(\mathrm{E2}:0_1^+ \to 2_2^+)$ are rather scattered:
$1.83(9)~e^2 \mathrm{fm}^4$, 
$1.6(2)~e^2 \mathrm{fm}^4$ in $(\alpha,\alpha^\prime)$ reaction \cite{It11}, 
$2.5(5)~e^2 \mathrm{fm}^4$ in $(\alpha,\alpha^\prime)$ reaction \cite{Jo03} as cited in Ref. \cite{Ko11}, 
3.7(65)~$e^2\mathrm{fm}^4$ from $(\gamma,\alpha)$ reaction Ref. \cite{Zi13}, and 
$14(+1.5,-2.0)~e^2\mathrm{fm}^4$ in reanalysis of the data  \cite{Zi13}, which is cited in Ref. \cite{Fr18}. % 10.025(50) 1.60(13)
Considering a large  model dependence of  $B(\mathrm{E2}:0_1^+ \to 2_2^+)$, more precise experimental infomation of this strength is desirable.

Although  the peaks corresponding to $2_4^+$ state in the E2-TSD for the AB(D) + $3\alpha$P models are clear, those for  the AB(A') + $3\alpha$P models are not so clear due to  broader widths.
This peak may correspond to the $2^+$ resonance  state:  $(E_r, \Gamma_{3\alpha})=$ [8.17(40) MeV, 1.77(20) MeV] reported in Ref. \cite{Ke17} with  uncertainty as shown in Table \ref{tab:12C-levels}.

%%%%%%%%%%%%%%%%%%%%%%%%%%%%%%
%----- FIGURE 3 -------------------------------------------------------
%%%%%%%%%%%%%%%%%%%%%%%%%%%%%%%%%
%---------------------------------------------------------------------

%%%%%%%%%%%%%%%%%%%%%%%%%%%
%%% Table VII:   2_2^+  2_3^+   2_4^+
%%%%%%%%%%%%%%%%%%%%%%%%%%%%
\begin{table*}[t]
\caption{
Calculated values of the resonance energy and the $3\alpha$-decay width for $2_2^+$, $2_3^+$, and $2_4^+$ states, and  the E2 transition strength for  $2_2^+$ state.
\label{tab:C12_2_energies} 
}
\begin{ruledtabular}
\begin{tabular}{cccccccc}
     & $E_r(2_2^+)$ & $\Gamma_{3\alpha}(2_2^+)$ &  $B(\mathrm{E2}:0_1^+ \to 2_2^+)$\footnote{Evaluated by integrating the E2-TSD for $E < 3.5$ MeV.}   & $E(2_3^+)$  & $\Gamma_{3\alpha}(2_3^+)$  & $E_r(2_4^+)$  & $\Gamma_{3\alpha}(2_4^+)$ \\
 Model    &      MeV    &  MeV   &  $e^2\mathrm{fm}^4$  & MeV   & MeV  & MeV & MeV \\
\hline
AB(A') + $3\alpha$P \\
A1 &   2.0 & 0.9  & 27.1 &   5.1 & 1.5    &   7.9  &  4.2      \\
A2 &   2.2 &   1.0  & 15.0&     5.1 &  1.8   &  8.7   &   5.2    \\
A3 &   2.2  &   1.0 &11.5 &    5.0 &   1.8 &   9.2  &    5.8   \\
A4 &   2.2 &   1.0  & 13.2&   4.8 &  1.5   &  9.4   &  6.5   \\
AB(D) + $3\alpha$P \\
D1 & 2.3 & 0.9  &13.5 &  4.6  & 0.2   & 6.6  & 1.8  \\
D2 & 2.5 & 0.9 &7.1  & 4.5  & 0.2  & 6.9  & 1.8 \\
D3 & 2.5 & 0.8 &4.9 &  4.5  & 0.2   & 7.1  & 1.8 \\
D4 & 2.5 & 0.9 & 3.1 &4.4 &  0.2 & 7.2  &  1.9  \\
%D5 & 2.6 & 0.9 &2.2 & 4.3 &  0.2 & 7.2  & 2.0   \\
\end{tabular}
\end{ruledtabular}
\end{table*}
%---------------------------------------------------------------------------

%%%%%%%%%%%%%%%%%
\subsection{$3\alpha(1^-)$ and $3\alpha(3^-)$ states}
%%%%%%%%%%%%%%%%%

\begin{figure}[tb]
\includegraphics[width=1.0\columnwidth]{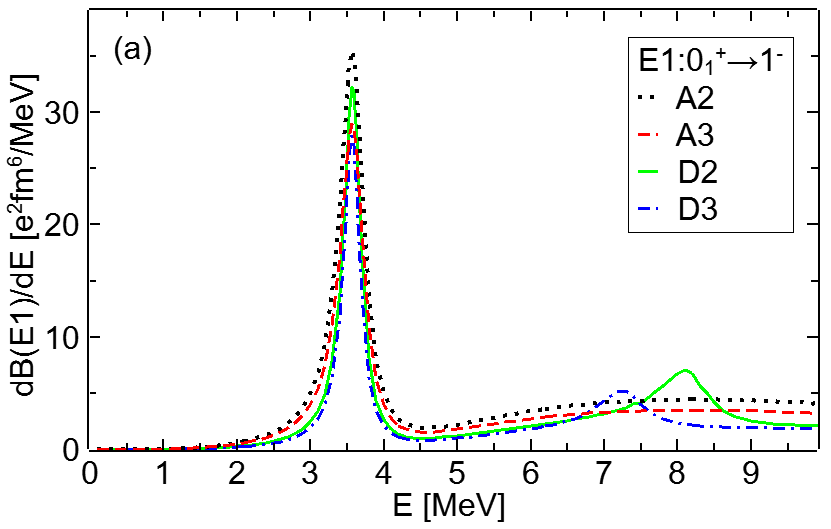}
\includegraphics[width=1.0\columnwidth]{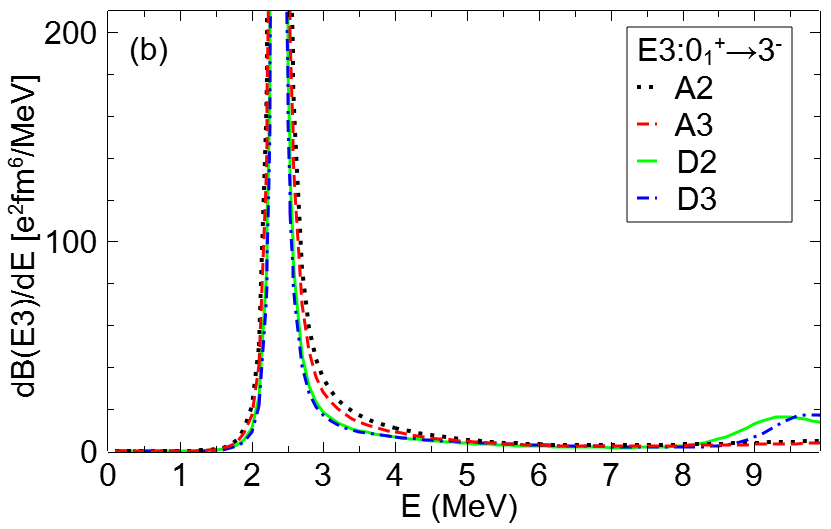}
\caption{(Color online) 
(a) The E1-TSDs  for the transition from ${}^{12}$C${(0_1^+)}$ to $3\alpha (1^-)$  state, 
and (b) the E3-TSDs  for the transition from ${}^{12}$C${(0_1^+)}$ to  $3\alpha (3^-)$ states.  
Dotted (black), dashed (red), solid (green), and dot-dashed (blue) curves  denote calculations with the A2, A3,  D2, and  D3 models, respectively.
\label{fig:db-de-1-3}
}
\end{figure}

Since there is no bound state for $3\alpha(1^-)$ and $3\alpha(3^-)$ states, the $3\alpha$P strength parameters for these states in Table \ref{tab:W3_data} are determined to reproduce the energy of the lowest resonance state shown in Table \ref{tab:12C-levels} by the energy of the first maximum in calculated TSD.
Calculated E1- and E3-TSDs for some selected models are shown in Fig. \ref{fig:db-de-1-3}. 

The $3\alpha$-decay width and  the E$\lambda$ strength calculated by fitting these TSDs to the resonance formula Eq.  (\ref{eq:dBdE-BW}) are shown in Tables \ref{tab:C12_1_results} and \ref{tab:C12_3_results} for  $3\alpha(1^-)$ and $3\alpha(3^-)$ states, respectively.

%%%%%%%%%%%%%%%%%%%%%%%
\subsubsection{${}^{12}\rm{C}(1_1^-)$ state}
%%%%%%%%%%%%%%%%%%%%%%%%

Calculated values of $\Gamma_{3\alpha}(1_1^-)$ for the AB(D) + $3\alpha$P models are consistent with the experimental value, but those for  the AB(A') + $3\alpha$P models are about 40\% larger than the experimental value.
The correlation between  $\Gamma_{3\alpha}(1_1^-)$ and the $3\alpha$P range parameter with the experimental value of $\Gamma_{3\alpha}(1_1^-)$ gives $a_3 = 3.1$ fm for the AB(D) + $3\alpha$P models.
% $a_3 = 3.13 \pm 0.13$ fm for the AB(D) + $3\alpha$P models.

Although the dependence of the E1 strength on  the $3\alpha$P range parameter is remarkable, there is no experimental information for these values in Ref. \cite{Ke17}. 

For the AB(D) + $3\alpha$P models, there appears another $1^-$ resonance peak at $E \sim 7$ MeV to 8 MeV, but there has not been reported   corresponding resonance state to this in the compilations so far \cite{Aj85,Aj90,Ke17}.

%%%%%%%%%%%%%%%%%%%%%%%
\subsubsection{${}^{12}\rm{C}(3_1^-)$ state}
%%%%%%%%%%%%%%%%%%%%%%%%

Calculated values of $\Gamma_{3\alpha}(3_1^-)$ have a weak dependence on the choice of the $3\alpha$P, but are determined essentially by  $2\alpha$P:  
about 55 keV for the AB(A') + $3\alpha$P calculations and 32 keV for the AB(D) + $3\alpha$P calculations.
The adapted value  in the compilation  \cite{Ke17},  $\Gamma_{3\alpha}(3_1^-)=46(3)$ keV, was recently criticized in Ref. \cite{Li24} to be updated by a smaller value, 38(2) keV. 
One of the grounds of this argument is related to the definition of the total decay width in the R-matrix parameterization. 
In the analysis of the R-matrix, an energy-dependent total decay width, which is referred to as a formal decay width (see Eq. (3) of Ref. \cite{Li24}), is highly model (in this case the channel radius parameter $a$) dependent and should not used for an intrinsic decay width, which is approximated by the observed  total width, Eq. (8) in Ref. \cite{Li24}. 
Actually, when the E3-TSF of the D3 model is fitted by the R-matrix parametrization with $a=4.0$ fm,  the formal decay width at the resonance peak energy becomes 48 keV, and the observed total width becomes 32 keV. 
Also, the formal (observed) decay width with $a=5$ fm becomes 36 keV (32 keV).
Thus the observed total width is model-independent and well agrees with the total width calculated from the line shape of the E3-TSF by Eq. (\ref{eq:dBdE-BW}). %,. which is referred to as the physical total width $\Gamma_{\mathrm{FWHM}}$ in Ref. \cite{Li24}.
Thus the values of $\Gamma_{3\alpha}(3_1^-)$ in Table \ref{tab:C12_3_results} should be compared to a smaller value, e.g., $38 (2)$ keV, not $46(3)$ keV. 

Calculated values of  the $B(\mathrm{E}3:0_1^+ \to 3_1^-)$ strength are largely model dependent.
However, experimental values are also widely distributed as follows:
In the compilation literature \cite{Ki02}, the value $6.1(0.9)\times10^2 ~e^2 \mathrm{fm}^6$ is adapted taking values obtained from inelastic  $(e,e^\prime)$ scattering: $7.5(1.0)\times10^2 ~e^2 \mathrm{fm}^6$  \cite{Cr67} and $4.75(1.20)\times10^2 ~e^2 \mathrm{fm}^6$ \cite{Cr66,Gu78}.
On the other hand,  values of $2.4(4)\times10^2 ~e^2 \mathrm{fm}^6$ \cite{Jo03} and $2.51(10)\times10^2 ~e^2 \mathrm{fm}^6$  \cite{It11} were obtained from inelastic  $(\alpha,\alpha^\prime)$ scattering.

For the AB(D) + $3\alpha$P  models, there appears another $3^-$ resonance peak for $E > 9$ MeV.
This peak could correspond to the $3^-$ resonance  state:  $(E_r, \Gamma_{3\alpha})=$ [11.3(1) MeV, 0.3 MeV] reported in Refs. \cite{Aj85,Aj90,Ke17}   as shown in Table \ref{tab:12C-levels}, although its isospin is not specified.

%%%%%%%%%%%%%%%%%%%%%%%%%%%%%%%
%----- FIGURE 4 -------------------------------------------------------
%%%%%%%%%%%%%%%%%%%%%%%%%%%%%%%
%---------------------------------------------------------------------

%%%%%%%%%%%%%%%%%%%%%%%%%%%%%%%%%%%
%%% Table VIII:   1_1^- 
%%%%%%%%%%%%%%%%%%%%%%%%%%%%%%%%%%%%%%%
\begin{table}[t]
\caption{
Calculated values of decay widths and E1  transition strengths for the $3\alpha(1_1^-)$ state.
\label{tab:C12_1_results}
 }
\begin{ruledtabular}
\begin{tabular}{ccc}
% $E_r(3_1^-) $ (MeV)    2.366
      &   $\Gamma_{3\alpha}(1_1^-)$ &  $B(\mathrm{E1}:0_1^+ \to 1_1^-)$ \\%  & $\Gamma_{\gamma}(1_1^-)$ \\
Model     &    keV    &  $e^2\mathrm{fm}^6$   \\%  & meV     \\
\hline
Exp. &  273(5) \footnote{Ref. \cite{Ke17}.}  &\\% &   \\
AB(A') + $3\alpha$P \\
A1    & 412 & 30 \\%& 1.363  \\
A2    & 392 & 21\\% & 0.887     \\
A3    & 383 & 16 \\%& 0.710     \\
A4    & 382 & 15\\% & 0.692       \\
AB(D) + $3\alpha$P \\
D1     & 297 & 21 \\%&0.840    \\  
D2     & 282 & 14\\% & 0.582  \\
D3     & 268 & 12\\% & 0.478 \\
D4     & 251 &9.0\\% & 0.367 \\ 
%D5     & 240 &7.0\\% & 0.299 \\
\end{tabular}
\end{ruledtabular}
\end{table}
%%%%%%%%%%%%%%%%%%%%%%%%%%%%%%%%%%%%%%%%%%%%%%%%%

%%%%%%%%%%%%%%%%%%%%%%%%%%%%%%%%%%%
%%% Table IX:   3_1^-
%%%%%%%%%%%%%%%%%%%%%%%%%%%%%%%%%%%%%%%
\begin{table}[t]
\caption{
Calculated values of decay widths and E3  transition strengths for the $3\alpha(3_1^-)$ state.
\label{tab:C12_3_results}
 }
\begin{ruledtabular}
\begin{tabular}{ccc}
% $E_r(3_1^-) $ (MeV)    2.366
      &    $\Gamma_{3\alpha}(3_1^-)$ &  $B(\mathrm{E3}:0_1^+ \to 3_1^-)$  \\% & $\Gamma_{\gamma}(3_1^-)$  &  $\Gamma_{\gamma}(3_1^-)/\Gamma_{3\alpha}(3_1^-) $ \\
Model     &   keV  &   $e^2\mathrm{fm}^6$  \\% &  meV   & $10^{-9}$ \\
\hline
Exp. &  46(3) \footnote{Ref. \cite{Ke17}.} & \\
      &      38(2) \footnote{Ref. \cite{Li24}.} \\%& \\
AB(A') + $3\alpha$P \\
A1    &  58 & 895 \\% &0.375 &   $6.5 $   \\
A2    & 56 & 638 \\% & 0.266 &    $ 4.8$    \\
A3    &  56 & 525 \\% & 0.218 &     $ 3.9$      \\
A4    &  55 & 491 \\% & 0.206 &    $ 3.7 $       \\
AB(D) + $3\alpha$P \\
D1     &  30 &  753 \\%& 0.316 & $11 $    \\  
D2     &  31 & 595 \\% &  0.247 & $8.0 $\\
D3     & 32& 503 \\%&  0.209 & $6.6 $\\
D4     &  32 & 388 \\%&  0.161& $5.0$\\ 
%D5     &   32 & 320\\% &  0.133 &$4.1 \times10^{-9}$\\
\end{tabular}
\end{ruledtabular}
\end{table}
%%%%%%%%%%%%%%%%%%%%%%%%%%%%%%%%%%%%%%%%%%%%%%%%%%%%%%%

%%%%%%%%%%%%%%%%%%
\section{Excitation-energy spectrum}
\label{sec:energy_spectrum}
%%%%%%%%%%%%%%%%%%%

In general, experimental information of the electric multipole transition strength distributions studied in the previous section
% for ${}^{12}\mathrm{C}(0_1^+)$ to $3\alpha$ states 
is included in cross sections of reactions leading to $3\alpha$ continuum states in the final state.
In this section, I will study the excitation-energy spectra in the $\alpha$-induced inelastic scattering ${}^{12}\mathrm{C}(\alpha,\alpha^\prime)3\alpha$ at $E_{\alpha,\mathrm{Lab}}=386$ MeV \cite{It11}. 
Calculations will be performed in PWIA, where the incident $\alpha$-particle collides with an $\alpha$-particle in the target, and is scattered to a certain solid angle $\Omega$  leaving $3\alpha$ with the excitation energy $E_x$ in the laboratory (Lab.) frame.
In this framework,  the cross section is  written in a factorized  form, 
\begin{equation}
\frac{d^2\sigma}{d\Omega dE_x}= N_K \sigma_{\alpha\alpha} R(E,Q), 
\label{eq:dsigma_PWIA}
\end{equation}
where $Q$ is the momentum transfer. 
A kinematical factor $N_K$ and  some other related kinematical variables are given in Appendix \ref{sec:kinematics}. 
The variable  $\sigma_{\alpha\alpha}$ is the cross section for the elementary process: $\alpha$-$\alpha$ scattering at $E_{\alpha,\mathrm{Lab}} = 386$ MeV. 
Since there is no available experimental data for the $\alpha$-$\alpha$ elastic scattering at this energy region,  in the present calculations, $\sigma_{\alpha\alpha}$  is taken to be the $\alpha$-$\alpha$ Coulomb (Rutherford) differential cross section as function of the momentum transfer  $Q$,  
\begin{equation}
\sigma_{\alpha\alpha}=\sigma_{C} (Q) = \left(\frac{\mu_{\alpha\alpha} Z_1  Z_2 e^2}{\hbar^2}\right)^2  \frac{4}{Q^4}, 
\end{equation}
where $\mu_{\alpha\alpha}$ is the reduced mass of $2\alpha$ system and $Z_1=Z_2=2$.
The factor $R(E,Q)$ is the response function corresponding to the transition from the $3\alpha$ bound state $\vert \Psi_b  (J_i M_i) \rangle$ with angular momentum $J_i$ and its third component $M_i$  to $3\alpha$-continuum states with energy $E$, 
\begin{equation}
R(E,Q) = \frac{1}{2J_i+1} \sum_{J_i M_i}\int d\bm{q}^\prime d\bm{p}^\prime  \delta\left(E-E^\prime\right)
\left\vert \langle \Psi^{(-)}_{\bm{q}^\prime, \bm{p}^\prime}(E^\prime)  \vert  
\sum_{i=1,3} e^{i \bm{Q}\cdot\bm{r}_i}
\vert \Psi_b(J_i M_i) \rangle \right\vert^2 ,
\label{eq:R_E}
\end{equation}
where  $\bm{Q}$ is  the momentum transfer vector. 
By taking the lowest order term in each of the multipole ($\lambda$) expansion of $e^{i \bm{Q}\cdot\bm{r}_i}$ up to $\lambda=3$, one has 
\begin{equation}
R(E,Q) = \sum_{\lambda=0,3}  a_\lambda(Q) \frac{1}{4e^2} \frac{dB(\mathrm{E}\lambda:J_i \to J_f)}{dE} , 
\end{equation}
where
\begin{equation}
a_\lambda (Q)  = 
{(4\pi)^2} \times 
\left\{
\begin{array}{cc}
( Q^2/6)^2/(4\pi) \quad\quad & (\mathrm{for}~ \lambda=0)
\\
(Q^3/30)^2  \quad\quad & (\mathrm{for}~  \lambda=1)
\\
(Q^2/15)^2  \quad\quad & (\mathrm{for}~  \lambda=2) 
\\
(Q^3/105)^3  \quad\quad & (\mathrm{for}~  \lambda=3).
\end{array}
\right. 
\label{eq:a_cfs}
\end{equation}

In this work, theoretical the cross section for the ${}^{12}\mathrm{C}(\alpha,\alpha^\prime)3\alpha$ reaction is expressed in the following form:
\begin{equation}
\frac{d^2\sigma}{d\Omega dE_x}= N_K  \frac{\sigma_{\alpha\alpha}}{4e^2} 
 \sum_{\lambda=0,3}  x_\lambda a_\lambda(Q)
\frac{dB(\mathrm{E}\lambda:J_i \to J_f)}{dE}, 
\label{eq:dcs-fit-formular}
\end{equation}
where the extra parameters $x_\lambda~ (\lambda=0,...,3)$ are introduced to take into account of various effects that are not incorporated in PWIA. 

To compare calculations with the data, possible experimental resolutions are taken into account by convoluting  Eq. (\ref{eq:dcs-fit-formular}) with a Gaussian function of  a full width at half maximum (FWHM). 
Thus fitting parameters in this work are weight factors of multipole components, $x_\lambda~(\lambda=0,1,2,3)$ in Eq. (\ref{eq:dcs-fit-formular}) and the convolution FWHM. 
The FWHM was searched for values between 100 keV and 250 keV with a step 10 keV  because the value of 200 keV was  quoted as the resolution in Ref. \cite{It11}. 
These  parameters are determined to reproduce  the experimental data in Ref. \cite{It11} for $8~ \mathrm{MeV} < E_x < 12~ \mathrm{MeV}$. 

For the  ${}^{12}\mathrm{C}(\alpha,\alpha^\prime)$ reaction at $E_{\alpha,\mathrm{Lab}}=386$ MeV with  $\theta=0^\circ$, the momentum transfer $Q$ takes 0.09 $\mathrm{fm}^{-1}$ at $E_x=8$ MeV  and 0.14 $\mathrm{fm}^{-1}$ at  $E_x=12$ MeV, and $N_K \approx 17.6$ through out the excitation energy range.
% $Q$ takes 18 MeV/c    28 MeV/c

\begin{figure}[tb]
\includegraphics[width=0.6\columnwidth]{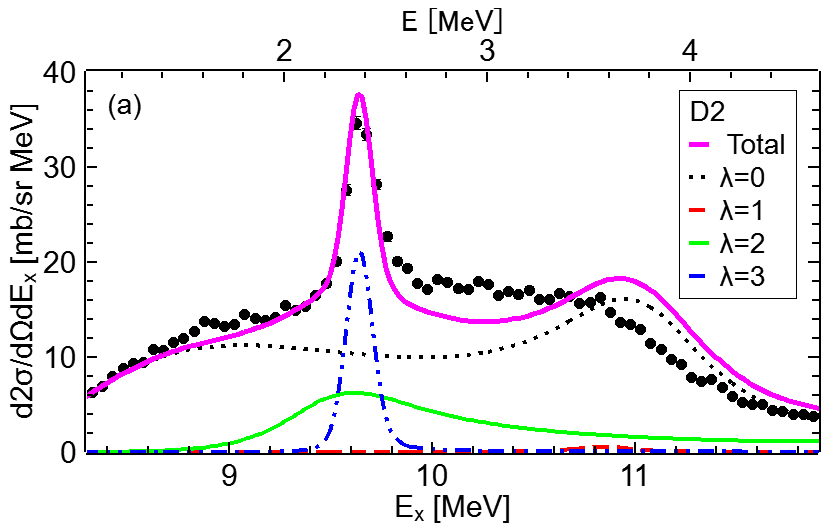}
\includegraphics[width=0.6\columnwidth]{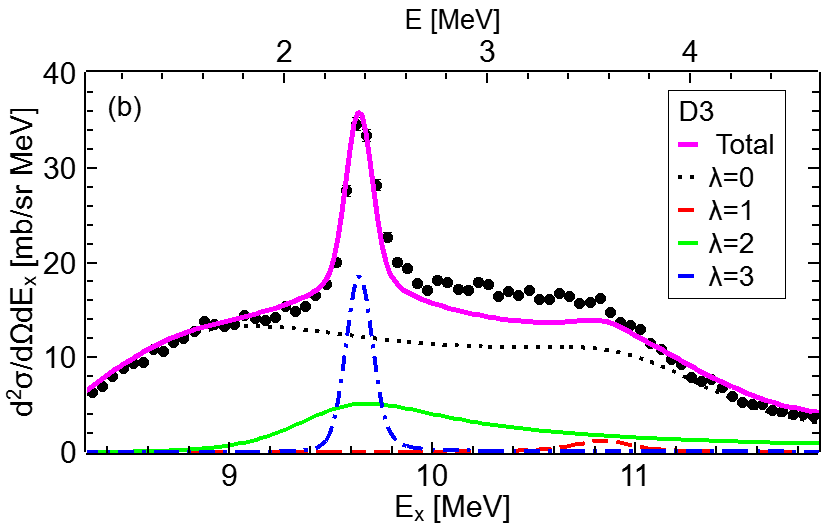}
\includegraphics[width=0.6\columnwidth]{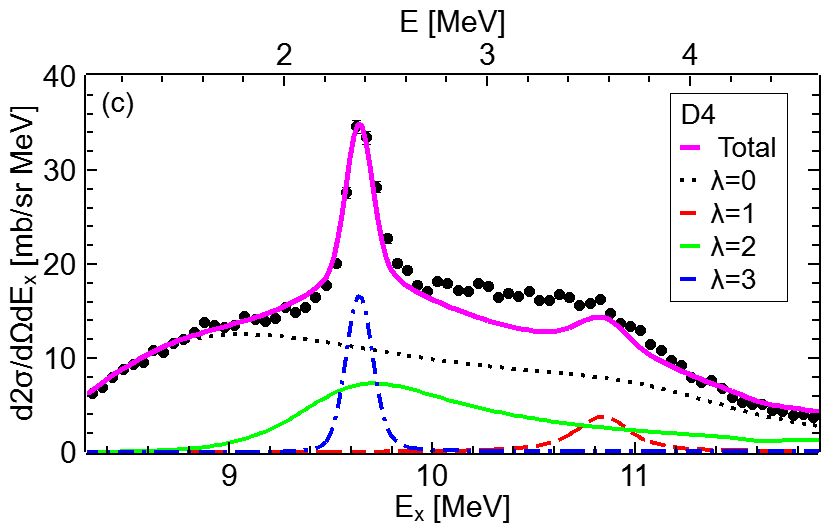}
%,angle=-90
\caption{(Color online) 
Energy spectra for the ${}^{12}\mathrm{C}(\alpha,\alpha^\prime)$ reaction at scattering angle $\theta_{\mathrm{lab}}=0^\circ$ for (a) D2, (b) D3, and (c) D4 models.
Bold solid (magenta), dotted (black), dashed (red), solid (green), and dot-dashed (blue) curves denote total, $\lambda=0$, 1, 2, and 3 contributions, respectively. 
Experimental data are taken from Ref. \cite{It11}.
\label{fig:386-0deg}
}
\end{figure}

It turns out that models other than the D2, D3, and D4 are not adequate to fit the experimental data with good quality mainly because of a large strength of $0_4^+$ compared to $0_3^+$ as explained below. 
The convolution FWHM=140 keV is best to obtain a good fit.
Fitted parameters $x_\lambda~(\lambda=0, \dots, 3)$ are shown in Table \ref{tab:PWIA_parameters}, and results of the spectrum are displayed in Fig. \ref{fig:386-0deg}.  
The spectrum consists of a sharp peak around $E_x = 9.6$ MeV, which corresponds to $3_1^-$ state, 
on the top of a broad bump contributed from $0^+$, $1^-$, and $2^+$ states. 

Although the $x_\lambda$-parameter search was performed automatically, the results can be interpreted as follows.
The lower part of the broad bump, $E_x < 9$ MeV, mainly consists of the $0_3^+$ state, and the parameter $x_0$ is determined to fit the spectrum in this region.
Then, the spectrum around the sharp peak consists of $2^+$ and $3^-$ states as well as the fixed $0^+$ state, from which the parameters $x_2$ and $x_3$ are determined.
Finally, the rest parameter $x_1$ is determined to reproduce the upper part of the broad peak, $E_x> 10.5$ MeV.

For D2, since the contribution of the $0_4^+$ peak is large and even exceeds the experimental data for $E_x > 11$ MeV, there is no room for the $1_1^-$ state. 
%The spectrum for $E_x > 10.5$ MeV comes from the peak of the  $0_4^+$ state and the high energy tail of the $2_2^+$ state for D2, but non-negligible contribution from $1_1^-$ peak for D3.
On the other hand, for D3 and D4, because of not too large contributions of the $0_4^+$ state, a room for the $1_1^-$ state together with the high energy tail of the $2_2^+$ peak, is left.
%a better fit for $E < 11$ MeV is obtained with   and thus . 
But it is remarked that the discrepancies between calculations and the data around  $E=10.5$ MeV are left even for these models.

For all of these models, the peak at $E_x \approx 9.7$ MeV consists of  the $3_1^-$ state together with rather broad $2_2^+$ peak and $0^+$ contributions.

The other models than the D2, D3, and D4, if the spectrum around $E_x=9$ MeV is fitted by the $0_3^+$ peak, contributions from the $0_4^+$ peak are too large to get a good fit.

%%%%%%%%%%%%%%%%%%%%%%%%%%%%%%%%%%%
%%% Table X:   PWIA parameters x0, x1. x2, x3,
%%%%%%%%%%%%%%%%%%%%%%%%%%%%%%%%%%%%%%%
\begin{table}[t]
\caption{
Parameters to fit the experimental energy spectra for the ${}^{12}\mathrm{C}(\alpha,\alpha^\prime)$ reaction at scattering angle $\theta_{\mathrm{lab}}=0^\circ$ \cite{It11} for the D2, D3, and D4 models.
\label{tab:PWIA_parameters}
 }
\begin{ruledtabular}
\begin{tabular}{ccccc}
Model      &   $x_0$ &  $x_1$   & $x_2$ &   $x_3$  \\
\hline
%$\theta=0^{\circ}$ \\
D2     & 0.957 & 0.509 & 0.102 & 2.66 \\
D3     & 1.15 & 1.34 & 0.128 & 2.78 \\
D4     & 1.59 & 5.12 & 0.295 & 3.23 \\ 
\end{tabular}
\end{ruledtabular}
\end{table}
%%%%%%%%%%%%%%%%%%%%%%%%%%%%%%%%%%%%%%%%%%%%%%%%%%%%%%%%%%%%%%%%%%%%

%%%%%%%%%%%%%%%%%%%%%%%%%%%%%%
%----- FIGURE 5 -------------------------------------------------------
%%%%%%%%%%%%%%%%%%%%%%%%%%%%%%
%---------------------------------------------------------------------

For the measurement of ${}^{12}\mathrm{C}(\alpha,\alpha^\prime)$ reaction at $E_{\alpha,\mathrm{Lab}}=386$ MeV with $\theta=3.7^\circ$, the momentum transfer $Q$ takes about 0.6 $\mathrm{fm}^{-1}$ and $N_K \approx 17.6$ through out the energy range considered in this work.

\begin{figure}[tb]
\includegraphics[width=1.0\columnwidth]{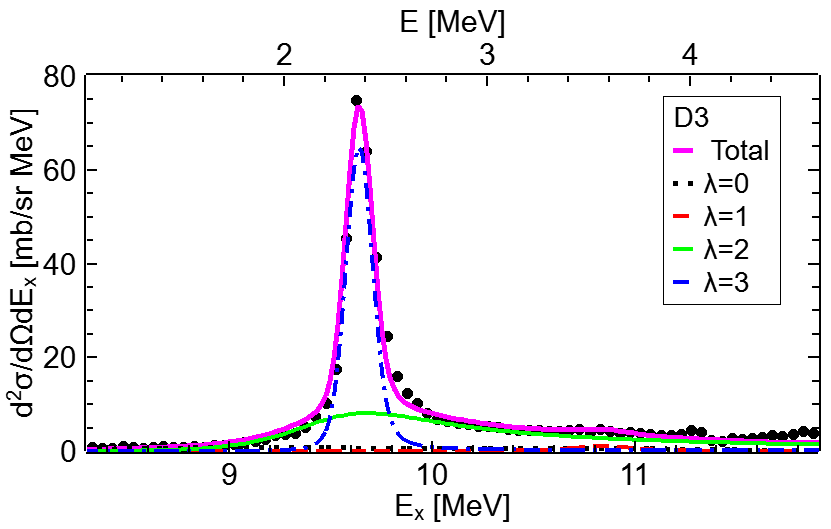} %,angle=-90
\caption{(Color online) 
Energy spectra for the ${}^{12}\mathrm{C}(\alpha,\alpha^\prime)$ reaction at scattering angle $\theta_{\mathrm{lab}}=3.7^\circ$ for the D3 model.
Magenta, black, red, green, and blue curves denote total, $\lambda=0$, 1, 2, and 3 contributions, respectively. 
Experimental data are taken from Ref. \cite{It11}.
\label{fig:386-37deg}
}
\end{figure}

The cross section for the measurement of ${}^{12}\mathrm{C}(\alpha,\alpha^\prime)$ reaction at $E_{\alpha,\mathrm{Lab}}=386$ MeV with  $\theta=3.7^\circ$ is shown in  Fig. \ref{fig:386-37deg} for the D3 model.
In this case, the contribution from $\lambda=3$ TSD dominates the spectrum and that from $\lambda=2$ TSD adds some component and other $\lambda=0$ and $\lambda=1$ contributions are small.
In the D3 model, fitted parameters are: $x_0=0.0692$, $x_1=0.0579$, $x_2=0.201$, and $x_3= 0.362$ with  FWHM=140 keV.

The existence of a peak corresponding $2_3^+$ state, which is predicted in the $3\alpha$ model, is not clear, because the cross sections for this state is too small and  smeared out by the convolution.

%%%%%%%%%%%%%%%%%%%%%%%%%%%%%%
%----- FIGURE 6 -------------------------------------------------------
%%%%%%%%%%%%%%%%%%%%%%%%%%%%%%
%---------------------------------------------------------------------

\begin{figure}[tb]
\includegraphics[width=1.0\columnwidth]{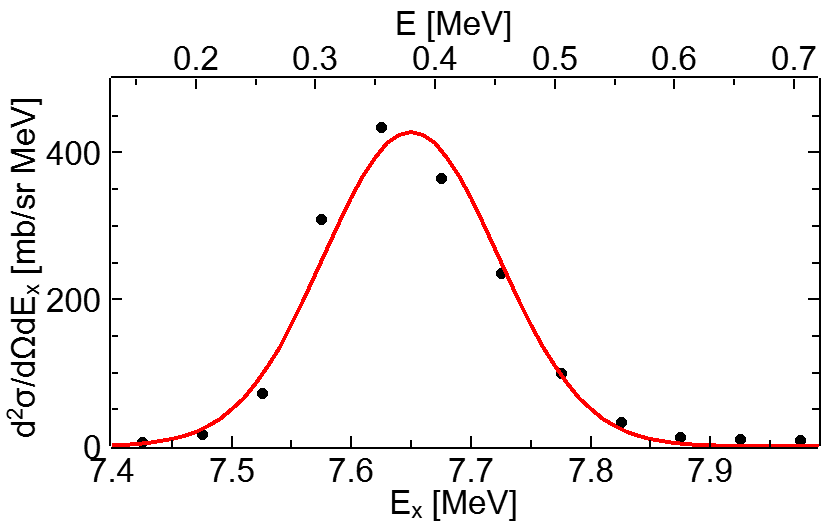}
\caption{(Color online) 
Energy spectra for the ${}^{12}\mathrm{C}(\alpha,\alpha^\prime)$ reaction at scattering angle $\theta_{\mathrm{lab}}=0^\circ$.
Red curve is the result of Gaussian fit with Eq. (\ref{eq:Gauss-fit-Hoyle}).
Experimental data are taken from Ref. \cite{It11}.
\label{fig:386-0deg-hoyle}
}
\end{figure}

Finally, I will analyze the cross section data around the Hoyle state.
Experimental data of the energy spectrum at $\theta_{\mathrm{Lab}}=0^\circ$ for $7.4~ \mathrm{MeV} < E_{x} < 7.9~ \mathrm{MeV}$ are fitted by a Gaussian distribution 
\begin{equation}
\frac{d^2\sigma}{d\Omega dE_x} = \left( \frac{N_K  \sigma_{\alpha\alpha}a_0}{4}\right) 
\frac{x_0 B_0}{e^2}
 \frac{1}{\sigma  \sqrt{2\pi}} \exp\left( -\frac{(E_x-E_r)^2}{2\sigma^2}   \right), 
\label{eq:Gauss-fit-Hoyle}
\end{equation}
where   $E_r=7.65$ MeV and $N_k \sigma_{\alpha\alpha}a_0/4=4.750$ $\mathrm{fm}^{-4}$ mb/sr.
The experimental data is well fitted with  $x_0 B_0=16.4 ~ e^2 \mathrm{fm}^4$ and  $\sigma  =  0.0724$ MeV (FWHM =170 keV) as shown in Fig. \ref{fig:386-0deg-hoyle}. 

This fitted value of $x_0 B_0$ is about 60\% of those calculated from the fitted values for the spectra of higher excitation-energy in Table \ref{tab:PWIA_parameters} together with the values of $B(\mathrm{E0}:0_1^+ \to 0_2^+)$  in Table \ref{tab:Hoyle-state}:  $x_0 B_0 = (27.7, 28.5, 28.0) ~ e^2 \mathrm{fm}^4$ for the D2, D3, and D4 models, respectively. 
In other words, the use of the parameters determined for resonance states at higher energies results the Hoyle state peak being overestimated.  
This is known as  the missing monopole strength of the Hoyle state \cite{Ko08}, which may occur due to the use of PWIA in the present calculations. 
It is suggested that  adequate treatments of distorting waves in initial and final states are necessary to resolve this problem \cite{Mi16}. 

%%%%%%%%%%%%%%%%%%%%%%%%%%%%%%
%----- FIGURE 7 -------------------------------------------------------
%%%%%%%%%%%%%%%%%%%%%%%%%%%%%%%
%---------------------------------------------------------------------

%%%%%%%%%%%%%
\section{Summary}
\label{sec:summary}
%%%%%%%%%%%%%%%

In this paper, low-energy continuum states as well as bound states in the ${}^{12}$C nucleus are studied by the $3\alpha$ models, where several sets of $2\alpha$- and $3\alpha$ potential models are introduced. 
The electric multipole strength distributions for the transition of ${}^{12}\mathrm{C}(0_1^+)$ state to $3\alpha$ continuum states with $J^\pi=0^+, 1^-, 2^+$, and $3^-$ are calculated, from which model dependence is studied for various observables such as the $3\alpha$-decay width and the electric multipole strength of $3\alpha$ resonance states.  
Using calculated strength distributions, the excitation energy spectra of the forward ${}^{12}\mathrm{C}(\alpha,\alpha^\prime)3\alpha$  reaction at $E_{\alpha,\mathrm{Lab}}=386$ MeV are calculated, and compared with the available experimental data. 

Two different versions of the $2\alpha$P, AB(A') and AB(D),  are used in the present calculations, which are two-range Gaussian-type along with the $\alpha$-$\alpha$ Coulomb potential. 
The $3\alpha$ potential models are constructed assuming some values of the range parameters with Gaussian form,  and thus the strength parameters are determined to reproduce the energies: $E_r(0_2^+)$, $E_r(1_1^-)$, $E(2_1^+)$, and $E_r(3_1^-)$. 

Of several models examined, the combination of AB(D) and $3\alpha$P with the range parameters $a_3=3.0$ fm (D3) can be recommended although there are some observables to be improved.  
Main reason of this choice comes from the ratio of the $0_3^+$ strength to $0_4^+$ strength,  which is important to fit the line shape of ${}^{12}\mathrm{C}(\alpha,\alpha^\prime)3\alpha$ energy spectrum.
In general, a longer range interaction, either $2\alpha$P or $3\alpha$P, results a larger  $0_4^+$ strength. 

There remain some rooms to improve interaction models, e.g., 
to reproduce the energies $E(0_1^+)$ and $E(0_2^+)$ simultaneously. 

Lastly, some states in $2^+$, $1^-$ and $3^-$ are observed in the calculations although they are not experimentally confirmed yet possibly because of a weak coupling to the ${}^{12}$C ground state and/or a large decay width.
For confirming the reliability of the $3\alpha$ model, the existence of these states are quite interesting. 

%%%%%%%%%%%%%%%%%%%
\begin{acknowledgments}
Numerical calculations in this work were performed using computational resources of the Laboratory provided by the
Research Center for Computing and Multimedia Studies, Hosei University (Project ID: LAB-667).
\end{acknowledgments}

%%%%%%%%%%%%%%%%%%%%%%%%%%%%%%%%%%%%%%%
\appendix

%%%%%%%%%%%%%
\section{Coulomb modified Faddeev equations }
\label{sec:Coul_mod_Faddeev}
%%%%%%%%%%%

In this section, a brief explanation of how to treat the long range Coulomb force in solving  Eq. (\ref{eq:Psi_def}) will be given.

For simplicity, $\alpha$-particles are considered to be interacting with two-body potentials only as  
\begin{equation}
\hat{V} = \sum_{i=1,3}V_i(x_i) =\sum_{i=1,3} \left[ V^\mathrm{S}(x_i) + \frac{(2e)^2}{x_i} \right], 
%\label{eq:aa-pot2}
\end{equation} 
where $V^\mathrm{S}(x)$ is the short range part of the interaction, and in this section I  use sets of Jacobi coordinates $\{\bm{x}_i, \bm{y}_i\}$,   
\begin{equation}
\bm{x}_i =  \bm{r}_j - \bm{r}_k, 
\quad\quad
\bm{y}_i  =  \bm{r}_i    - \frac12 \left(\bm{r}_j + \bm{r}_k \right) 
\label{eq:Jacobi_2}
\end{equation}
with $(i,j,k)$ being $(1,2,3)$ or its cyclic permutations.

I start with the differential equation form of Eq. (\ref{eq:Psi_def}),
\begin{equation}
\left(  E -  \hat{H}_{3\alpha} \right) \vert \Xi \rangle =  \hat{O} \vert \Psi_b \rangle.
\label{eq:diff_Psi}
\end{equation}
In the Faddeev theory \cite{Fa61}, the three-body wave function is decomposed into three Faddeev components, 
\begin{equation}
\vert \Xi \rangle = \vert\Phi^{(1)} \rangle+ \vert\Phi^{(2)} \rangle + \vert\Phi^{(3)}\rangle, 
\label{eq:Fad-dec}
\end{equation}
and coupled equations of these components (Faddeev equations) describe the multiple scattering of three particles with rearranging the interacting pair.
In the case of short range forces, $\vert\Phi^{(i)}\rangle$ describes a process in which the pair particles $(j,k)$ are interacting while the particle $i$ is free as the spectator.
However, in the case of $3\alpha$ system, this picture is not adequate because the spectator $i$ is not free due to the long range Coulomb forces.

To remedy this problem, Coulomb-modified Faddeev equations are introduced as follows \cite{Sa79,Is13}:
\begin{eqnarray}
 \left[ E - H_0 - V_1(x_1) - \frac{2 (2 e)^2}{y_1}  \right] \vert \Phi^{(1)} \rangle   
&=& \hat{O}_{1}  \vert \Psi_b \rangle 
+  \left[V_1(x_1)  - \frac{(2 e)^2}{y_2}  \right]  \vert \Phi^{(2)} \rangle  
\cr 
&& 
+ \left[ V_1(x_1) - \frac{(2 e)^2}{y_3}  \right] \vert\Phi^{(3)}  \rangle, 
\cr
\mathrm{(and ~cyclic~ permutations)} ,&&
\label{eq:Fad-eq-C}
\end{eqnarray}
where  the transition operator is decomposed into three components,
\begin{equation}
\hat{O} = \hat{O}_{1} + \hat{O}_{2} + \hat{O}_{3},     
\label{eq:Hgam-dec}
\end{equation}
with the condition that $\hat{O}_{i}$ is symmetric with respect to the exchange of $j$ and $k$.
It is stressed that the original  equation (\ref{eq:diff_Psi}) is obtained by summing up all equations in (\ref{eq:Fad-eq-C}). 

The auxiliary potential $\frac{2 (2 e)^2}{y_i}$ in the left hand side of Eq. (\ref{eq:Fad-eq-C}) plays a role to introduce a Coulomb distortion of the spectator $i$. 
In the right hand side, the auxiliary potentials,  $\frac{(2 e)^2}{y_2}$ and  $\frac{(2 e)^2}{y_3}$, play a role to cancel the long range components in the potentials $V_1(x_1)$, namely $\frac{(2 e)^2}{x_1}$, which makes  the potential terms
$(2 e)^2\left( \frac1{x_1} - \frac1{y_2} \right)$  and  $(2 e)^2\left( \frac1{x_1} - \frac1{y_3} \right)$, 
short range functions with respect to the variable $x_1$.
While this cancellation holds sufficiently for bound states and continuum states below three-body breakup threshold \cite{Is03}, it does insufficiently for the case of  the three-body breakup reaction \cite{Is09}. 
To avoid difficulties arising from this, a mandatory cutoff factor $e^{-(x/R_C)^4}$ is introduced.
This is an approximation made for this calculations, and $R_C=35$ fm is used in the calculations, which was found to be enough to obtain  converged results \cite{Is13}.

%%%%%%%%%%%%%
\section{Kinematical relations in  ${}^{12}\mathrm{C}(\alpha,\alpha^\prime)3\alpha$ reaction}
\label{sec:kinematics}
%%%%%%%%%%%

In this appendix, some kinematical notations and relations to describe the ${}^{12}\mathrm{C}(\alpha,\alpha^\prime)3\alpha$ reaction treated relativistically in partial are summarized.

Let $T_{i}$ ($T_f$) and $\bm{K}_{i}$ ($\bm{K}_{f}$) be the kinetic energy and momentum (wave number), respectively, of  the incident $\alpha$-particle (outgoing $\alpha$-particle) in Lab. frame.
Here, 
\begin{eqnarray}
K_{i} &=& \sqrt{ \left( m_\alpha c^2 + T_i \right)^2 - m_\alpha^2 c^4 } /( \hbar c)
\\
K_{f} &=& \sqrt{ \left( m_\alpha c^2 + T_f \right)^2 - m_\alpha^2 c^4 }/( \hbar c) .
\end{eqnarray}

The total energy $E_{\mathrm{tot}}^{(L)}$,  energy transfer $\omega^{(L)}$, and momentum transfer $\bm{Q}^{(L)}$ in Lab. frame are 
\begin{eqnarray}
E_{\mathrm{tot}}^{(L)} &=&  m_\alpha c^2 + T_i + m_{\mathrm{C}} c^2,
\\
\omega^{(L)} &=&  T_i  -  T_f, 
\\
\bm{Q}^{(L)} &=& \bm{K}_{i} - \bm{K}_{f},
\end{eqnarray}
where  $m_{\mathrm{C}}$ is the mass of the target ${}^{12}$C nucleus.

The total energy $E_{tot}^{(c)}$,  energy $E_{i}^{(c)}$ and momentum $k_i^{(c)}$ of the incident $\alpha$-particle, energy $E_{\mathrm{C}}^{(c)}$ of the target  in the initial $\alpha$-${}^{12}\mathrm{C}$ ($\alpha$-C)  c.m. frame are
\begin{eqnarray}
E_{tot}^{(c)}  &=& \sqrt{E_{tot}^{(L) 2} - (\hbar c K_{i})^2},
\\
E_{i}^{(c)} &=& \frac{E_{tot}^{(c) 2} + m_\alpha^2 c^4 - m_{\mathrm{C}}^2  c^4}{2 E_{tot}^{(c)}},
\\
k_i^{(c)} &=&  \sqrt{E_{i}^{(c) 2}-m_\alpha^2 c^4},
\\
E_{\mathrm{C}}^{(c)} 
& =&  \frac{E_{tot}^{(c) 2} - m_\alpha^2  c^4 + m_{\mathrm{C}}^2  c^4}{2 E_{tot}^{(c)} }.
\end{eqnarray}

The energy $E_{f}^{(c)}$ and momentum $k_f^{(c)}$ of the outgoing $\alpha$-particle, and  
energy $E_{3\alpha}^{(c)}$ of the $3\alpha$-system in the final $\alpha$-$3\alpha$ c.m. frame are
\begin{eqnarray}
E_{f}^{(c)} &=& \frac{E_{tot}^{(c) 2} + m_\alpha^2  c^4 - (E_{3\alpha}^{(L) 2}- ( \hbar c Q^{(L)})^2)}{2 E_{tot}^{(c)} },
%\end{equation}
\\
k_f^{(c)} &=& \sqrt{E_{f}^{(c) 2} - m_\alpha^2  c^4} / (\hbar c).
\\
E_{3\alpha}^{(c)} 
&=& \frac{E_{tot}^{(c) 2} - m_\alpha^2  c^4 + (E_{3\alpha}^{(L) 2}- (\hbar c Q^{(L)} )^2 )}{2 E_{tot}^{(c)} },
\end{eqnarray}
where
\begin{equation}
E_{3\alpha}^{(L)} =  m_{\mathrm{C}} c^2 + \omega^{(L)}.
\end{equation}

The energy transfer $\omega^{(c)}$ and momentum transfer $\bm{Q}^{(c)}$ in the  $\alpha$-C c.m. frame are 
\begin{eqnarray}
\omega^{(c)} &=& E_{i}^{(c)} - E_{f}^{(c)}
\\
\bm{Q}^{(c)} &=& \bm{k}_i^{(c)} - \bm{k}_f^{(c)}, 
\end{eqnarray}
and $Q$ in Eq. (\ref{eq:dsigma_PWIA}) is defined  as $Q=\vert \bm{Q}^{(c)}  \vert$.

The energy of the $3\alpha$-system $E$ in the  $3\alpha$ c.m. frame is 
\begin{equation}
E = \sqrt{E_{3\alpha}^{(c) 2} - (\hbar c k_f^{(c)})^2} - 3 m_\alpha c^2.
\label{Eq:E_3p}
\end{equation}

The kinematical factor $N_K$ in Eq. (\ref{eq:dsigma_PWIA}) is 
\begin{equation}
N_{K} = 
\frac{\mu_i \mu_f}{\mu_{\alpha\alpha}^2} \frac{K_f}{k_i^{(c)}} \frac{d\Omega^{(c)}}{d\Omega}
 \frac{dE}{d\omega^{(c)}},
\label{eq:N_K}
\end{equation}
where  $\Omega^{(c)}$ is the solid angle of the ejected $\alpha$-particle in the  $\alpha$-C c.m. frame;  $\mu_i$ and $\mu_f$ are
\begin{eqnarray} 
\mu_i c^2 &=& \frac{E_{i}^{(c)} E_{\mathrm{C}}^{(c)}}{E_{i}^{(c)} + E_{\mathrm{C}}^{(c)}},
\\
\mu_f c^2 &=& \frac{E_{f}^{(c)} E_{3\alpha}^{(c)}}{E_{f}^{(c)} + E_{3\alpha}^{(c)}},
\end{eqnarray}
and
\begin{equation}
\frac{dE}{d\omega^{(c)}} 
= \frac{E_{tot}^{(c)}}{E + 3 m_\alpha c^2}.
\label{eq:dE_domega_cm}
\end{equation}

%%%%%%%%%%%%%%%%%%
%%% References %%%
%%%%%%%%%%%%%%%%%%

\end{document}